\UseRawInputEncoding
\documentclass[aps,pre,reprint,10pt,superscriptaddress,showpacs]{revtex4-2}
\usepackage{amsfonts} 
\usepackage{amsmath}
\usepackage{amssymb}
\usepackage{graphicx}
\usepackage{subfigure}
\usepackage{color}
\usepackage{color}
\usepackage{soul}
\usepackage{cancel}
\usepackage{ulem}

\begin{document}
\title{Transverse instability of electron-acoustic solitons in a relativistic degenerate
astrophysical magnetoplasma}
\author{A. P. Misra}
\email{apmisra@visva-bharati.ac.in}
\homepage{Author to whom any correspondence should be addressed.}
\affiliation{Department of Mathematics, Siksha Bhavana, Visva-Bharati University, Santiniketan 731 235, West Bengal, India}
\author{A. Abdikian}
\email{abdykian@gmail.com}
\affiliation{Department of Physics, Malayer University, Malayer 65719-95863, Iran}
\date{\today}

\begin{abstract}
We study the nonlinear theory of small-amplitude electron-acoustic solitons (EASs) in a relativistic astrophysical magnetoplasma consisting of two-temperature electrons: a sparse population of relativistic nondegenerate classical electrons and a group of fully degenerate dense relativistic electrons (main constituent) immersed in a static magnetic field with a neutralizing stationary ion background. By using the multiple-scale reductive perturbation technique with the Lorentz transformation, the Zakharov-Kuznetsov (ZK) and the modified Zakharov-Kuznetsov (mZK) equations are derived to describe the evolution of EASs in two different regimes of relativistic degeneracy: $r_{d0}<50$ and $r_{d0}\gtrsim50$, where $r_{d0}=\left(n_{d0}/n_{\rm{cr}}\right)^{1/3}$ is the degeneracy parameter in which $n_{d0}$ is the unperturbed number density of degenerate electrons and $n_{\rm{cr}}\approx6\times10^{29}~\rm{cm}^{-3}$ the critical number density at which $p_F=m_ec$ and which defines the emergence of the relativistic regime. Here, $p_F$ is the Fermi momentum, $m_e$ is the invariant electron mass, and $c$ is the speed of light in vacuum. The characteristics of the plane soliton solutions of ZK and mZK equations and the soliton energy are studied.  We show that the solitons moving at an angle $\alpha$ to the external magnetic field can be unstable under transverse long-wavelength perturbations.  The growth rates of instabilities are obtained and analyzed with the effects of the relativity parameter $\beta_{\rm{cl}}=k_BT_{\rm{cl}}/m_ec^2$ and the degeneracy parameter $r_{d0}$, where $k_B$ is the Boltzmann constant and $T_{\rm{cl}}$ is the temperature of classical electrons.  Interestingly, the ZK solitons, even if it is stable for the first-order perturbations, can be unstable in the second-order correction. Furthermore, while the first-order growth rates of perturbations for ZK solitons tend to vanish  as $\alpha\rightarrow 38^\circ$, that for the mZK soliton goes to zero as $\alpha\rightarrow 90^\circ$.  However, depending on the angle $\alpha$, the growth rates are found to be reduced either by increasing the values of $\beta_{\rm{cl}}$ or by decreasing the values of $r_{d0}$. The applications of our results to astrophysical plasmas, such as those in the environments of white dwarfs are discussed.  

\end{abstract}
\maketitle
\section{Introduction}\label{intro}
In many equilibrium plasmas, such as those driven by external sources like beams or waves, electrons can be grouped in different species (with different densities) according to their different thermal energies. Such a division is also possible in low-temperature plasmas where electrons may not be in thermal equilibrium due to, e.g., their collisions with neutral atoms, strong electromagnetic fields, and a large spatial gradient field. Typically, such plasmas are known as two-electron-temperature plasmas or plasmas with two-temperature electrons. The latter was first experimentally observed by Sheridan \textit{et al.} \cite{sheridan1991} more than three decades back in a sputtering magnetron plasma. Later, such plasmas were produced in various plasma environments, including the recent one by Sharma \textit{et al.} \cite{sharma2022} in a low-pressure glow discharge plasma.
\par
Plasmas with two groups of electrons can play essential roles in laboratory systems, e.g., they influence the properties of a sheath in plasma processing and support coherent nonlinear structures like solitons or shocks \cite{sharma2022}. They also appear in high-density degenerate plasmas such as those in laser-produced or beam-driven plasmas and compact astrophysical objects like white dwarfs and neutron stars where the system energy flows mainly into the electrons and electrons can have a relatively high-energy tail: $m_ec^2<k_BT<k_BT_F\equiv m_ec^2(G_d-1)$, where $m_e$ is the invariant electron rest mass, $c$ is the speed of light in vacuum, $k_B$ is the Boltzmann constant,  $T~(T_F)$ is the thermodynamic (Fermi) temperature of electrons, and $G_d=\sqrt{1+r_d^2}$ is the relativistic factor for degenerate electrons with $r_d\equiv p_F/m_ec$ denoting the degeneracy parameter and  $p_F\equiv m_e c\left(n_d/n_{\rm{cr}}\right)^{1/3}$ the Fermi momentum. Here, $n_{\rm{cr}}=m^3_ec^3/3\pi^2\hbar^3\approx6\times10^{29}~\rm{cm}^{-3}$ is the critical number density at which $p_F=m_ec$ and which defines the emergence of the relativistic regime. The electron background density thus deviates from the thermodynamic equilibrium, resulting in a group of sparse population (low-density) of classical (non-degenerate) relativistic electrons and a highly degenerate dense relativistic electron gas. Such exciting and unusual states of matter may coexist, e.g., during the formation of relativistic jets in the magnetosphere of accretion-induced collapsing white dwarfs \cite{kryvdyk2007}. In such extreme environments, both electron species can achieve the speed of light in vacuum $(c)$. Also,  the number density of degenerate electrons is so high (which usually exceeds the critical density $n_{\rm{cr}}$  even for the minor compact objects, white dwarfs) that the Fermi energy $(k_BT_F)$  dominates over the thermal energy $(k_BT)$ in the relevant dynamics, i.e., $k_BT\ll k_BT_F\equiv m_ec^2\left(G_d-1\right)$, where  $G_d\gg1$ corresponds to highly relativistic degenerate systems with $p_F\gg m_e c$ or, $r_d\gg1$. It has been argued that the radiation spectra coming from compact astrophysical objects \cite{negi1991,wood2006,ogelman1977} can carry essential information from the localization of electrostatic or electromagnetic waves (e.g., formation of solitons) in the core of these massive stars composed of degenerate electrons and non-degenerate ions. Furthermore, most of the compact astrophysical objects are under the influence of strong magnetic fields. Thus, magnetoplasmas with  stationary ions where both the classical and degenerate dense electrons can have relativistic speeds and the dense species can be of relativistic degeneracy are of great interest due to their potential applications in astrophysics.
\par
In $1977$, Watanabe and Taniuti \cite{watanabe1977} reported that plasmas with two-temperature electrons could excite a low-frequency mode,  called electron-acoustic wave (EAW) whose phase velocity lies in between the thermal counterparts of cold and hot electrons, if the number density of hot species is much higher than that of cold ones. In this context, several authors have focused on the dynamics of low-frequency EAWs owing to their relevance in laboratory, space, and astrophysical plasmas \cite{roy2022,shatashvili2020,pakzad2020,misra2021,li2024}. Recently, the nonlinear theory of large amplitude EAWs has been studied by Ali \textit{et al.} \cite{ali2023} in an unmagnetized two-electron-temperature plasma with relativistic degeneracy of cold electrons using a one-dimensional fluid model without considering the relativistic motion of fluid particles. However, as stated before, strong magnetic fields greatly influence most compact astrophysical objects, and particles emanating from different sources are not only highly degenerate but also can achieve the speed of light in vacuum. Thus, the relativistic treatment in the dynamics of degenerate electrons becomes necessary to study the nonlinear propagation of EAWs in magnetized degenerate two-electron-temperature plasmas.
\par 
It is to be noted that in  fully degenerate (at zero temperature) plasmas, typical collisions between two species are forbidden because of Pauli's exclusion principle. The latter drastically reduces the collision rate since all degenerate electrons have energies below the Fermi energy $E_F$, and no transition is possible because there are no available states for electrons to occupy. When the velocity of classical species $(u_{\rm{cl}})$ becomes smaller than the Fermi velocity of degenerate electrons $(v_F)$,  as the classical electron moves in the plasma, it slows down, giving its kinetic energy to the degenerate electrons. However, since the Fermi-sea gets occupied by degenerate electrons, only electrons on the Fermi-surface can collide with classical electrons. It has been shown that \cite{son2006}  if $u_{\rm{cl}}\ll v_F$ and the normalized Wigner-Seitz radius $r_s\equiv (3/4\pi n_d)^{1/3}/a\ll1$ (which holds for the number density of degenerate species, $n_d\gg10^{24}~\rm{cm}^{-3}$ and $a=\hbar^2/m_e e^2$ is the Bohr radius), the collision frequency (almost independent of the number density of degenerate electrons) is of the order of $\sim10^{13}/$ s, i.e., the collision time is $\tau_c\sim10^{-13}$ s corresponding to the particle density  $n_d\sim10^{28}~\rm{cm}^{-3}$. The reduction of the collision frequency is about $10-20\%$ when the relativistic degenarcy effects are pronounced \cite{son2006}. On the other hand, the time scale for electron-acoustic waves is $\tau_p\sim\omega^{-1}_{\rm{ecl}}\sim10^{-17}$ s (corresponding to the particle density $n_d\sim100n_{\rm{cl}}\sim10^{28}~\rm{cm}^{-3}$ or as mentioned above), which is much smaller than $\tau_c\sim10^{-13}$ s. Here, $\omega_{\rm{ecl}}$ denotes the plasma oscillation frequency of classical electrons. The time scale $\tau_p$ tends to get reduced (and remains much smaller than $\tau_c$) as the electron number density increases. Furthermore, since degenerate electrons have higher kinetic energies than classical electrons, and the collision rate with classical electrons is significantly low, they can travel greater distances at velocities approaching the speed of light in vacuum $c$. Thus, in the regimes of high-thermal motion $(\beta_{\rm{cl}}\equiv k_BT_{\rm{cl}}/m_ec^2>1)$ of sparse electrons and strong degeneracy $(r_d>1)$ of dense electrons and  as long as $u_{\rm{cl}}\ll v_F$ and $r_s\ll1$ for which $\tau_p\ll\tau_c$ holds, the two groups of electrons are separable. 
\par The objective of this work is to advance the previous theory of small-amplitude EAWs in multi-dimensional relativistic degenerate astrophysical magnetoplasmas whose main constituents are the bulk highly degenerate dense relativistic electrons mixed with a small concentration of classical nondegenerate electrons. Starting from a set of fully relativistic multi-dimensional fluid equations and using the perturbation expansion scheme with the Lorentz transformations relevant to the relativistic dynamics, we derive the Zakharaov-Kuznetsov (ZK) and modified ZK (mZK) equations, which govern the evolution of small-amplitude electron-acoustic solitons in two different regimes of relativistic degeneracy. We also study the stability of these solitons and show that they are unstable under transverse plane wave perturbations while propagating obliquely to the external magnetic field. The growth rates of instabilities can be suppressed by either changing the orientation of propagation or the parameters associated with the relativistic degeneracy and the relative thermal energy of classical electrons.
\section{The model}\label{sec-model}
We consider the nonlinear propagation of EAWs in a fully relativistic magnetized plasma consisting of two groups of electrons, namely the sparsely populated relativistic nondegenerate classical (cl) electrons   and dense population of relativistic fully degenerate (d) electrons (main component), and immobile singly charged positive ions (i). The plasma is immersed in a static magnetic field along the $z$-axis (${\bf B}=B_0\hat{z}$). At equilibrium, the quasineutrality condition reads
\begin{equation} \label{eq-neutra}
n_{\rm{cl0}}+n_{d0}=n_{i0},~ {\rm{i.e.,}}~\frac{n_{i0}}{n_{d0}}=1+\delta,  
\end{equation}
where $n_{j0}$ stands for the equilibrium number density of $j$-th species particles and $\delta={n_{\rm{cl0}}}/{n_{d0}}$ is the ratio between the equilibrium number densities of classical and degenerate electrons such that   $\delta\ll1$.
\par  
The dynamics of relativistic classical and degenerate electron fluids is given by (For details of the derivation of the flyuid model, see Appendix \ref{sec-appendix} and for similar models, see, e.g., Refs. \cite{shatashvili2020,misra2018,shatashvili2020})  
\begin{equation}\label{eq-cont}
 \frac{\partial \left(\gamma_jn_j\right)}{\partial t} +\nabla\cdot (\gamma_jn_j\textbf{u}_j)=0, 
 \end{equation}
\begin{equation}\label{eq-moment}
 \begin{split}
 \frac{\gamma_jH_j}{c^2}\frac{d}{dt} (\gamma_j {\bf u}_j)&=-en_j\gamma_j \left({\bf E}+\frac{1}{c}{\bf u}_j\times {\bf B} \right)\\
 &-\left(\nabla+\frac{\gamma_j^2{\bf u}_j}{c^2}\frac{d}{d t}\right)P_j,  
 \end{split}
 \end{equation}
  \begin{equation}\label{eq-poiss}
 \nabla^2 \phi=4\pi e (\gamma_{\rm{cl}}n_{\rm{cl}}+\gamma_dn_d-n_{i0}),  
 \end{equation}
 where $d/dt\equiv\partial/\partial t+ {\bf u}_j \cdot \nabla$ is the total derivative, $e$ is the elementary charge,   $\gamma_j=1/\sqrt{1-u_j^2/c^2}$ is the Lorentz factor for fluid motion; ${\bf E}$ and ${\bf B}$ are, respectively, the electric and magnetic fields; and  $n_j$,  ${\bf u}_j$, and $P_j$ are, respectively, the number density (proper), velocity, and pressure of classical ($j=\rm{cl}$) and degenerate ($j=d$) electrons. Also, $H_j$ is the enthalpy per unit volume of the electron fluids, given by, \cite{berezhiani1995} 
  \begin{equation}\label{eq-Hj}
  H_{j}=n_{j}m_ec^2G_{j},
\end{equation}    
where $G_j$ is the effective mass factor for electrons. For classical species, $G_{\rm{cl}}(\sigma_{\rm{cl}})=K_3(\sigma_{\rm{cl}})/K_2(\sigma_{\rm{cl}})$ with $K_2(\sigma_{\rm{cl}})$ and $K_3(\sigma_{\rm{cl}})$ denoting the MacDonald functions of the second and third orders respectively, and $\sigma_{\rm{cl}}=m_ec^2/k_BT_{\rm{cl}}=1/\beta_{\rm{cl}}$ the relativity parameter associated with the thermal energy of classical electrons. Here, $T_{\rm{cl}}$ is the invariant temperature of classical electrons. Also, for degenerate species, we have $G_d=\sqrt{1+r_d^2}$, where $r_d=p_F/m_ec=\left(n_d/n_{cr}\right)^{1/3}$ is the dimensionless parameter such that at equilibrium, the relativistic gamma factor is $\gamma_F\equiv G_{d0}=\sqrt{1+r_{d0}^2}$ with $r_{d0}=\left(n_{d0}/n_{\rm{cr}}\right)^{1/3}$ denoting the degeneracy parameter and $n_{\rm{cr}}$ the critical number density of degenerate electrons defined before.
Note that the degenerate electrons can be highly relativistic if $p_F\gg m_ec$ or $r_{d0}\gg1$, i.e., $n_{d0}\gg n_{\rm{cr}}$. In such dense degenerate environments, the Fermi energy $(k_BT_F)$ strongly dominates over the thermal energy $(k_BT)$ of degenerate species, i.e., $k_BT\ll k_BT_F=m_ec^2\left(\gamma_F-1\right)$. 
 \par 
 We mention that the factors $G_{\rm{cl}}$  and $G_d$  for the effective mass $(m_eG_j)$  are quite distinctive. While the former depends on the classical temperature $T_{\rm{cl}}$, the latter depends on the number density of degenerate electrons. For nonrelativistic thermal motion of classical electrons, we have $\beta_{\rm{cl}}\ll1$ for which the  effective mass reduces to $m_eG_{\rm{cl}}\sim m_e+5k_BT_{cl}/2c^2$, i.e., in the cold plasma limit, $G_{\rm{cl}}=1$, while in the limit of high-temperature (or ultra-relativistic) fluid flow,  $(\beta_{\rm{cl}}\gg1)$, the effective mass becomes $m_eG_{\rm{cl}}\sim 4k_BT_{cl}/c^2\gg m_e$, which gives $G_{\rm{cl}}=4\beta_{\rm{cl}}\gg1$. On the other hand, the limits, $r_{d0}\ll1$ and $r_{d0}\gg1$, respectively, correspond to the regimes of nonrelativistic (or weakly relativistic) and ultra-relativistic degenerate electron gas. 
\par 
In what follows, we consider the following equations of state for the classical \cite{berezhiani1995} and fully degenerate \cite{chandrasekhar1935} electrons.
\begin{equation}
 P_{\rm{cl}}=n_{\rm{cl}}k_BT_{\rm{cl}},\label{eq-Pcl} 
 \end{equation}
 \begin{equation}\label{eq-Pd}
  P_d=\frac{m_e^4c^5}{24\pi^2\hbar^3}\left[r_d\left(2r_d^2-3\right)\left(1+r_d^2\right)^{1/2}+3\sinh^{-1}r_d\right], 
 \end{equation}  
where $\hbar=h/2\pi$ is the reduced Planck's constant.
\par 
It is imperative to make the system of equations \eqref{eq-cont}-\eqref{eq-Pd} dimensionless by redefining the variables as follows:
\begin{equation}
\begin{split}
 &n_j\rightarrow\ n_j/n_{j0},~  {\bf{u}}_j\rightarrow{\bf{u}}_j/c,~\phi\rightarrow\ e\phi/m_ec^2, \\
 & (x,z)\rightarrow\left(\omega_{\rm{ecl}}/c\right)(x,z),~ t\rightarrow\omega_{\rm{ecl}}t,~\Omega=\omega_c/\omega_{\rm{ecl}},\\
 & P_j\rightarrow\ P_j/n_{j0}m_ec^2,~H_j\rightarrow\ H_j/n_{j0}m_ec^2, 
  \end{split} 
\end{equation}
where $\omega_{\rm{ecl}}=\sqrt{4\pi n_{\rm{cl0}}e^2/m_e}$ is the plasma oscillation frequency of classical electrons and $\omega_c=eB_0/m_ec$ is the electron cyclotron frequency. Thus, Eqs. \eqref{eq-cont}-\eqref{eq-poiss} reduce to
 \begin{equation}\label{eq-cont1}
 \frac{\partial \left(\gamma_jn_j\right)}{\partial t} +\nabla\cdot (\gamma_jn_j\textbf{u}_j)=0, 
 \end{equation}
\begin{equation}\label{eq-moment1}
 \begin{split}
 \gamma_jH_j\frac{d}{dt} (\gamma_j {\bf u}_j)&=n_j\gamma_j \left(\nabla\phi- \Omega{\bf u}_j\times \hat{z}  \right)\\
 &-\left(\nabla+\gamma_j^2{\bf u}_j\frac{d}{d t}\right)P_j,  
 \end{split}
 \end{equation}
  \begin{equation}\label{eq-poiss1}
 \nabla^2 \phi=\gamma_{\rm{cl}}n_{\rm{cl}}+\frac{1}{\delta}\gamma_dn_d-\left(1+\frac{1}{\delta}\right).  
 \end{equation}
\section{ZK equation and its soliton solution}\label{sec-ZK}
We study the nonlinear evolution of small-amplitude electron-acoustic solitons in relativistic degenerate dense magnetoplasmas. 
In the weakly nonlinear theory of two-dimensional relativistic wave dynamics, we employ the reductive perturbation technique to derive an evolution equation for the first-order perturbation of the electrostatic potential $\phi$. To this end, we use the Lorentz transformations and define a new frame of reference (that moves along the $z$-axis with the phase velocity $v_p$  of EAWs in the laboratory frame)  in which the space and time variables are stretched as \cite{lee2009}
\begin{equation}
\begin{split}
&X=\varepsilon^{1/2}x,~Z=\varepsilon^{1/2}\gamma_v\left(z-v_pt\right),\\
&T=\varepsilon^{3/2}\gamma_v\left(t-v_pz\right),\label{eq-stretch}
\end{split}
\end{equation}
where $\varepsilon$ is a small expansion parameter measuring the weakness of perturbations and $\gamma_v=1/\sqrt{1-v_p^2}$ is the Lorentz factor (in which $v_p$ is normalized by $c$) for the relativistic dynamics of EAWs. Here, we mention that while the use of the Galilean transformation can be reasonably good for the evolution of nonrelativistic perturbations \cite{williams2013}, the  Lorentz transformations should be considered for relativistic
 wave motions \cite{lee2009}. Later, we will see that the Lorentz factor $\gamma_v$ contributes to the wave dispersion and nonlinearity of the evolution equations of EAWs. 
Furthermore, the dependent variables are expanded as 
\begin{equation}
S=S^{(0)}+\varepsilon\ S^{(1)}+\varepsilon^2S^{(2)}+\varepsilon^3S^{(3)}+\cdots,\label{eq-expan}
\end{equation}
where $S^{\left(n\right)}$ denotes the $n$-th order perturbation and $S^{(0)}$ the equilibrium state, which is defined for the normalized dependent variables as: $\left(n_{cl}, n_d, u_{cl}, u_d, \phi\right)=\left(1,1,0,0,0\right)$.
\par 
 We substitute the stretched coordinates [Eq. \eqref{eq-stretch}] and the expansion \eqref{eq-expan} into Eqs. \eqref{eq-cont1}-\eqref{eq-poiss1}, and equate different powers of $\varepsilon$. In the lowest order of $\varepsilon$, we obtain the following linear dispersion relation for the wave phase velocity. 
\begin{equation} \label{eq-phase}
v_p=\left[\frac{3\beta_{\rm{cl}}R_0+\delta r_{d0}^2}{3R_0\left(G_{\rm{cl}}+\delta R_0^2\right)}\right]^{1/2},
\end{equation}
where $R_0=\sqrt{1+r_{d0}^2}$ is a factor associated with the degree of degeneracy of dense electron gas.
\begin{figure*}
	\centering
	\includegraphics[width=6in,height=3in]{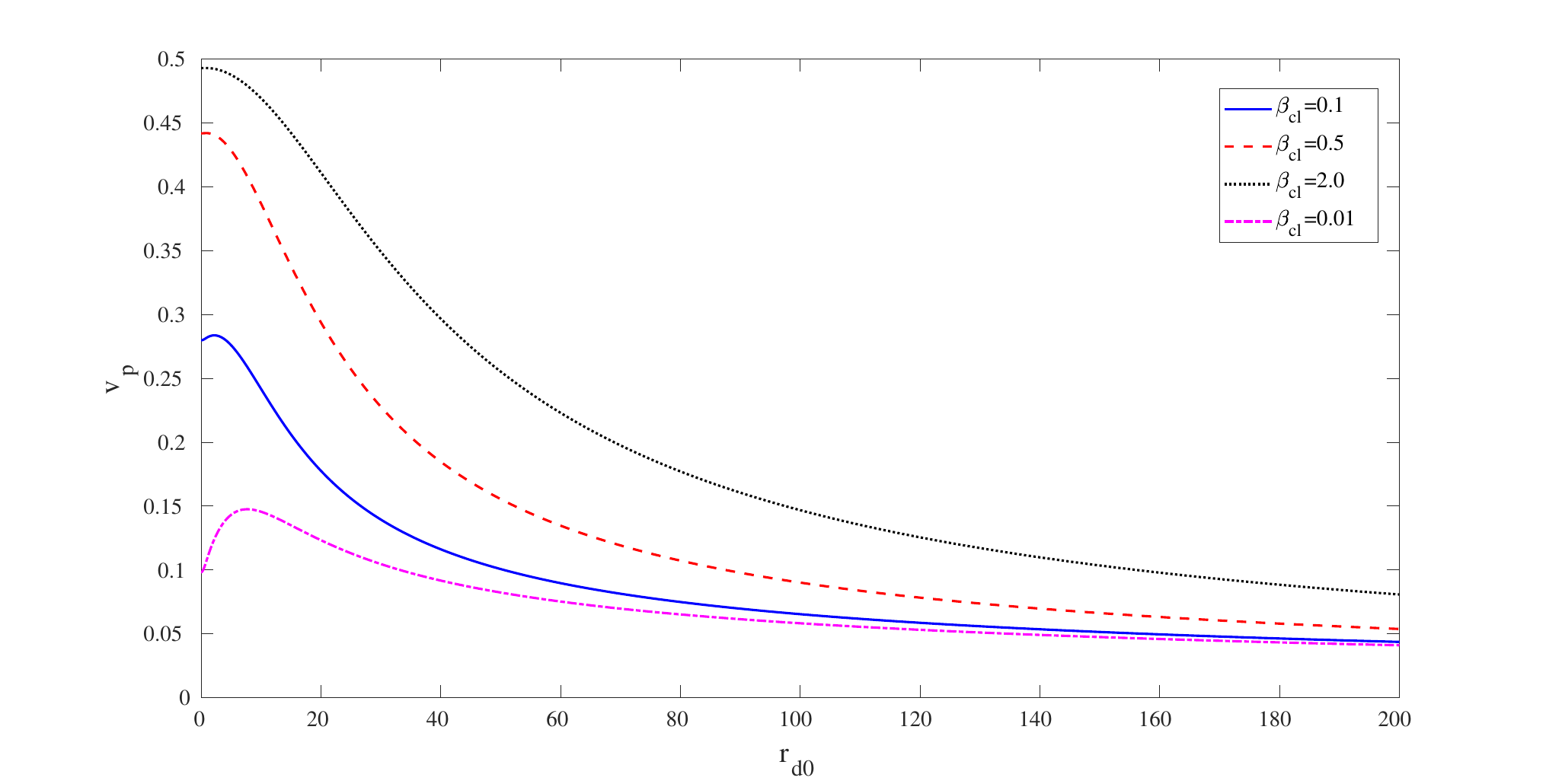}
	\caption{The phase speed ($v_p$) is plotted against the relativistic degeneracy parameter  ($r_{d0}$) for different values of the relativity parameter $\beta_{\rm{cl}}$ as in the legend  but with a fixed value of $\delta$: $\delta=0.01$.}  \label{fig1-phase}
\end{figure*}
From Eq. \eqref{eq-phase} it is clear that the phase velocity depends on three   parameters, namely, the number density ratio of classical to degenerate electrons $\delta$, the degeneracy parameter $r_{d0}$, and the relativity parameter  $\beta_{\rm{cl}}$. Basically, this phase velocity corresponds to that of long-wavelength electron-acoustic perturbations and the wave becomes dispersionless in the limit of small wave number, i.e., $k\rightarrow0$. Such an assertion can easily be verified from the linear dispersion relation of EAWs using the Fourier mode analysis of Eqs. \eqref{eq-cont1}-\eqref{eq-poiss1}. Since $\delta\ll1$, the contributions from the terms proportional to $\delta$ can be significant in the regime of ultra-relativistic degeneracy of dense electrons with  $r_{d0}\gg1$. In the latter, we have $v_p\approx r_{d0}/\sqrt{3R_0^3}$. However, in the weakly relativistic degeneracy regime $(r_{d0}\ll1)$, the contribution of $\delta$ is insignificant, and we have $v_p\approx\sqrt{\beta_{\rm{cl}}/G_{\rm{cl}}}$, i.e., the classical reult is recovered.
\par 
In the graphical plots, we will consider wide ranges of values of the dimensionless parameters, namely $\beta_{cl}$ and $r_{d0}$, to observe the characteristics of EAWs in the regimes of low-, moderate, or high-thermal motion of classical electrons and weakly, moderate, or ultra-relativistic degenerate regimes of dense electrons.  The dimensionless parameter values in the graphs may not be as arbitrary as they will appear. Still, they will correspond to some specific regimes of astrophysical settings, especially for the plasma densities and classical temperatures. For example, in Fig. 1, $\beta_{cl}=0.1$ corresponds to the temperature $T_{cl}\sim 10^8$ K.   Also, since $r_{d0}$ is related to the degenerate particle density, i.e., $r_{d0}=\left(n_{d0}/n_{cr}\right)^{1/3}$, one can easily calculate the particle number density for a given value of $r_{d0}$ and identify the corresponding degeneracy regime in astrophysical settings. Later, we will also discuss the ranges of values of the parameters in which the present theory is applicable in Secs. \ref{sec-applica} and \ref{sec-summary}.   
\par 
The characteristics of the phase velocity are shown graphically  in Fig. \ref{fig1-phase} against the parameter $r_{d0}$, ranging from weakly to ultra-relativistic degeneracy regimes, for different values of the thermal relativity parameter $\beta_{\rm{cl}}$. It is seen that the phase velocity decreases [except for the case of $\beta_{\rm{cl}}<1$, where the phase velocity initially increases in a small interval of $r_{d0}$ (See the solid and dash-dotted lines)] with increasing values of $r_{d0}$. However, it reaches a steady state value for $r_{d0}\gg1$. Thus, long-wavelength EAWs can propagate with a constant phase speed in ultra-relativistic degenerate plasmas.  
On the other hand, as the relativity parameter $\beta_{\rm{cl}}$ increases, including when the thermal energy of classical electrons exceeds the kinetic energy, the phase velocity increases, i.e., the crest of the wave travels faster but it remains smaller than $c$.  In the limits of nonrelativistic classical temperature, i.e., $\beta_{\rm{cl}}\ll1$ and nonrelativistic degeneracy with $r_{d0}\ll1$, we have $v_p^2\approx \beta_{\rm{cl}}/(1+\delta)$, which in original dimension, gives $v_p\sim v_t/\sqrt{1+\delta}<v_t$, where $v_t=\sqrt{k_BT_{\rm{cl}}/m_e}$ is the thermal velocity of classical electrons. In the high-temperature or ultra-relativistic limit, i.e.,    $\beta_{\rm{cl}}\gg1$  with nonrelativistic degeneracy $r_{d0}\ll1$, we have $v_p\approx0.5$, i.e., in original dimension, $v_p\sim c/2$. This is evident from Fig. \ref{fig1-phase} that as $\beta_{\rm{cl}}$ becomes larger and larger, $v_p$ approaches more or less a constant value $0.5$ (See the dotted line). It is to be noted that in a regime of $\beta_{\rm{cl}}>1$,  a small increment of the value of $\delta$ can significantly reduce the value of the phase velocity (not shown in the figure). Furthermore, it may be concluded that in contrast to nonrelativistic nondegenerate classical plasmas \cite{holloway1991}, the phase velocity of EAWs in relativistic degenerate plasmas can be smaller than the thermal velocity of classical electrons.  For example, for $\beta_{\rm{cl}}=0.1$, which corresponds to the temperature, $T_{cl}\sim6\times10^8$ K as in the interior of white dwarfs \cite{saumon2022,valyavin2011,kryvdyk2007,kryvdyk1999,misra2012},  we have the thermal velocity, $v_t\approx0.3$ and the phase velocity lies in the interval $0.05<v_p<0.3$ (see the solid line in Fig. \ref{fig1-phase}).   
\par 
Proceeding to consider the nonlinear perturbations, i.e., equating the coefficients of the next higher-order of $\varepsilon$, i.e., $\varepsilon^2$,  from Eqs. \eqref{eq-cont1} and \eqref{eq-moment1}, we obtain the following equations for classical and degenerate electrons.
\begin{eqnarray}
&&\frac{\partial n_d^{(1)}}{\partial T}-v_p\frac{\partial u_{dz}^{(1)}}{\partial T}+\frac{\partial\left(n_d^{(1)}u_{dz}^{(1)}\right)}{\partial Z}-v_p\frac{\partial n_d^{(2)}}{\partial Z}\nonumber\\
&&-v_pu_{dz}^{(1)}\frac{\partial u_{dz}^{(1)}}{\partial Z}+\frac{\partial u_{dz}^{(2)}}{\partial Z}+\frac{1}{\gamma_v}\frac{\partial u_{dx}^{(2)}}{\partial X}=0,\label{eq-nd1}
\end{eqnarray}
\begin{eqnarray}
&&\frac{\partial n_{cl}^{(1)}}{\partial T}-v_p\frac{\partial u_{clz}^{(1)}}{\partial T}+\frac{\partial\left(n_{cl}^{(1)}u_{clz}^{(1)}\right)}{\partial Z}-v_p\frac{\partial n_{cl}^{(2)}}{\partial Z}\nonumber\\
&&-v_pu_{clz}^{(1)}\frac{\partial u_{clz}^{(1)}}{\partial Z}+\frac{\partial u_{clz}^{(2)}}{\partial Z}+\frac{1}{\gamma_v}\frac{\partial u_{clx}^{(2)}}{\partial X}=0,\label{eq-ncl1}
\end{eqnarray}
\begin{eqnarray}
&&\Omega_eu_{dy}^{\left(3\right)}-R_0^2v_p\gamma_v\frac{\partial u_{dx}^{\left(2\right)}}{\partial Z}+R_0^2\gamma_vu_{dz}^{\left(1\right)}\frac{\partial u_{dz}^{\left(1\right)}}{\partial Z}-\frac{\partial\phi^{\left(2\right)}}{\partial X}\nonumber\\
&&+\frac{r_{d0}^2}{9R_0^3}\left(-R_2^2n_d^{\left(1\right)}\frac{\partial n_d^{\left(1\right)}}{\partial X}+3R_0^2\frac{\partial n_d^{\left(2\right)}}{\partial X}\right)=0,\label{eq-udy}
\end{eqnarray}
\begin{eqnarray}
\Omega_eu_{dx}^{(3)}+R_0^2v_p\gamma_v\frac{\partial u_{dy}^{(2)}}{\partial Z}=0,\label{eq-udx}
\end{eqnarray}
\begin{eqnarray}
&&-3R_0^3\frac{\partial u_{dz}^{\left(1\right)}}{\partial T}+r_{d0}^2v_p\frac{\partial n_d^{(1)}}{\partial T}-3R_0v_p\frac{\partial\phi^{(1)}}{\partial T}\nonumber\\
&&+n_d^{(1)}\biggl(\frac{1}{3} r_{d0}^2\frac{R_2^2}{R_0^2}\frac{\partial n_d^{\left(1\right)}}{\partial Z}+3R_0\left(3+5r_{d0}^2\right)v_p\frac{\partial u_{dz}^{\left(1\right)}}{\partial Z}\biggr)\nonumber\\
&&+\Biggl[u_{dz}^{\left(1\right)}\left(r_{d0}^2v_p\frac{\partial n_d^{\left(1\right)}}{\partial Z}-3R_0^3\frac{\partial u_{dz}^{\left(1\right)}}{\partial Z}\right)-r_{d0}^2\frac{\partial n_d^{\left(2\right)}}{\partial Z}\nonumber\\
&&+3R_0^3v_p\frac{\partial u_{dz}^{\left(2\right)}}{\partial Z}+3R_0\frac{\partial\phi^{\left(2\right)}}{\partial Z}\Biggr]=0,\label{eq-udz}
\end{eqnarray}
\begin{eqnarray}
&&\Omega_eu_{cly}^{(3)}-G_{cl}v_p\gamma_v\frac{\partial u_{clx}^{(2)}}{\partial Z}-\beta_{cl}n_d^{(1)}\frac{\partial n_{cl}^{(1)}}{\partial X}+\beta_{cl}\frac{\partial n_{cl}^{(2)}}{\partial X}\nonumber\\
&&-\frac{\partial\phi^{(2)}}{\partial X}=0,\label{eq-ucly}
\end{eqnarray}

\begin{eqnarray}
-\Omega_eu_{clx}^{(3)}-G_{cl}v_p\gamma_v\frac{\partial u_{cly}^{(2)}}{\partial Z}=0,\label{eq-uclx}
\end{eqnarray}

\begin{eqnarray}
&&\beta_{cl}v_p\frac{\partial n_{cl}^{(1)}}{\partial T}-G_{cl}\frac{\partial u_{clz}^{(1)}}{\partial T}-v_p\frac{\partial\phi^{(1)}}{\partial T}+\beta_{cl}n_d^{(1)}\frac{\partial n_{cl}^{(1)}}{\partial Z}\nonumber\\
&&+\beta_{cl}v_pu_{clz}^{(1)}\frac{\partial n_{cl}^{(1)}}{\partial Z}-\beta_{cl}\frac{\partial n_{cl}^{(2)}}{\partial Z}+G_{cl}v_pn_{cl}^{(1)}\frac{\partial u_{clz}^{(1)}}{\partial Z}\nonumber\\
&&-G_{cl}u_{clz}^{(1)}\frac{\partial u_{clz}^{(1)}}{\partial Z}+G_{cl}v_p\frac{\partial u_{clz}^{(2)}}{\partial Z}+\frac{\partial\phi^{(2)}}{\partial Z}=0,\label{eq-uclz}
\end{eqnarray}
where $R_2=\sqrt{1+2r_{d0}^2}$.
\par 
Similarly, by considering the terms in the coefficients of $\varepsilon^2$, we obtain  the following expression from the Poisson equation \eqref{eq-poiss1}.
  \begin{eqnarray}
&&n_{cl}^{\left(2\right)}+\frac{1}{\delta}n_d^{\left(2\right)}+\frac{1}{2}\left(u_{clz}^{\left(1\right)}\right)^2+\frac{1}{2\delta}\left(u_{dz}^{\left(1\right)}\right)^2\nonumber\\
&&-\gamma_v^2\frac{\partial^2\phi^{\left(1\right)}}{\partial Z^2}-\frac{\partial^2\phi^{\left(1\right)}}{\partial X^2}=0.\label{eq-pois2}
\end{eqnarray}
Finally, by eliminating the second-order perturbed quantities from Eqs. \eqref{eq-nd1}-\eqref{eq-pois2}, we obtain after a few steps the following  ZK equation for the evolution of small-amplitude electron-acoustic solitary waves in magnetized relativistic  degenerate plasma.   
\begin{eqnarray}
\frac{\partial\psi}{\partial T}+A_1\psi\frac{\partial\psi}{\partial Z}+A_2\frac{\partial^3\psi}{\partial Z^3}+A_3\frac{\partial^3\psi}{\partial Z\partial X^2}=0,\label{eq-ZK}
\end{eqnarray}
where $\psi=\phi^{(1)}$. The nonlinear $(A_1)$ and the dispersion coefficients ($A_2$ and $A_3$) are 
  are given by
\begin{equation}
\begin{split}
&A_1=-\frac{\gamma_v^2}{D}\biggl[\frac{\beta_{cl}\left(3\beta_{cl}R_0+2r_{d0}^2v_p^2-6R_0^3v_p^4\right)}{\left(\beta_{cl}-G_{cl}v_p^2\right)^3\left(r_{d0}^2-3R_0^3v_p^2\right)}\\
&+\frac{3r_{d0}^2R_2^2+9R_0^2\left(-9R_0^3+2r_{d0}^2+R_0\left(3+5r_{d0}^2\right)\right)v_p^2}{\left(r_{d0}^2-3R_0^3v_p^2\right)^3\delta}\\
&-\frac{G_{cl}v_p^2\left[2\left(r_{d0}^2-3R_0^3v_p^2\right)+{3\beta_{cl}R_0}\right]}{\left(\beta_{cl}-G_{cl}v_p^2\right)^3\left(r_{d0}^2-3R_0^3v_p^2\right)}\biggr],
\end{split}
\end{equation}
\begin{equation}
\begin{split}
&A_2=\frac{\gamma_v^4}{D},~A_3= \frac{\gamma_v^2}{D}\biggl[1   \\
& +\frac{v_p^4\left(G_{cl}^3\left(r_{d0}^2-3R_0^3v_p^2\right)-{3R_0^7\left(\beta_{cl}-G_{cl}v_p^2\right)}\right)}{\left(\beta_{cl}-G_{cl}v_p^2\right)^2\Omega_e^2\left(r_{d0}^2-3R_0^3v_p^2\right)}\biggr],\label{eqd25}
\end{split}
\end{equation}
with
\begin{eqnarray}
D=\frac{2v_p\left(G_{\rm{cl}}r_{d0}^2-3\beta_{cl}R_0^3\right)}{\left(\beta_{cl}-G_{\rm{cl}}v_p^2\right)^2\left(r_{d0}^2-3R_0^3v_p^2\right)}.\nonumber
\end{eqnarray}
It is evident that both the nonlinear and dispersion coefficients are significantly modified by the degeneracy parameter $r_{d0}$ (associated with the dense relativistic fully degenerate electron gas at zero temperature)  and the classical relativity parameter $\beta_{\rm{cl}}$                                                                         (associated with the sparse nondegenerate relativistic classical electrons). However, the external magnetic field only contributes to the wave dispersion coefficient $A_3$ and varies with it inversely, implying that a weak magnetic field strength can result into a strong wave dispersion leading to the  wave broadening.  It is interesting to note that the Lorentz factor $\gamma_v$ explicitly appears in both the dispersion and nonlinear coefficients and thus enhances their magnitudes, especially when the phase velocity of EAWs is no longer negligible compared to the speed of light in vacuum. This occurs especially in a regime of weakly or moderately relativistic degeneracy (which corresponds to the density regime, $n_{d0}\sim10^{28}-10^{30}~\rm{cm}^{-3}$) and where the electron thermal energy exceeds the rest mass energy (\textit{cf}. the dotted line in Fig. \ref{fig1-phase}). The latter may correspond  to a regime of high temperature exceeding the typical temperature ($\sim10^8$ K) in the core of white dwarfs before they start  to cool down. From the coefficients $A_1,~A_2$ and $A_3$, it is also noted that their magnitudes can be exceedingly high  when the phase velocity of EAWs approaches a value: $v_p\approx\sqrt{\beta_{{\rm{cl}}}/G_{\rm{cl}}}$ or $v_p\approx r_{d0}/\sqrt{3R_0^3}$, which, as mentioned before, corresponds to regimes of weakly or ultra-relativistic degeneracy of dense electrons. However, such higher values of the dispersion or nonlinear coefficients may not be permissible for the excitation of electron-acoustic solitons, as the soliton formation requires a delicate balance between the dispersion and nonlinear effects. So, in our analysis we will skip the discussion on these two particular limits. 
\par 
We study the evolution of electron-acoustic solitons in relativistic degenerate plasmas and see whether the solitary waves once formed can propagate with a speed larger than the linear phase speed of EAWs. Typically, a soliton is formed due to a delicate balance between the nonlinearity (which causes the wave steepening) and the wave dispersion (responsible for wave broadening). So, it is pertinent to study the characteristics of the dispersion and nonlinear coefficients of the ZK equation \eqref{eq-ZK}. However, before we investigate the behaviors of them, we first obtain a stationary soliton solution of Eq. \eqref{eq-ZK}. To this end, we recast Eq. \eqref{eq-ZK} in the following form
\begin{equation}
\begin{split}
\frac{\partial\psi^\prime}{\partial T^\prime}+\psi^\prime\frac{\partial\psi^\prime}{\partial Z^\prime}&+\sigma_{2}\frac{\partial^3 \psi^\prime}{\partial^2Z^{\prime3}}\\
&+\sigma_{3}\frac{\partial^3 \psi^\prime}{\partial Z^{\prime}\partial^2X^{\prime2}}=0,\label{eq-ZK-nond1}
\end{split}
\end{equation}
where $\psi^\prime=\sigma_{1}|A_1|\psi$, $T^\prime=T/\sqrt{|A_2|}$,  $X^\prime=X/\sqrt{|A_3|}$, $Z^\prime=Z/\sqrt{|A_2|}$, and $\sigma_j=\rm{sgn}(A_j)$, i.e., $\sigma_j=\pm1$  according to when $A_j\gtrless0$, $j=1,2,3$.  
For traveling wave solution, we apply  the transformation $\xi^\prime=l^\prime_xX^\prime+l^\prime_zZ^\prime-u^\prime_0T^\prime$, 
 where $l^\prime_z=\cos\alpha$ and $l^\prime_x=\sin\alpha$ are the direction cosines along the $x$- and $z$-axes, and   $u^\prime_0$ is the constant speed of the moving frame of reference, which corresponds to a small increment in the phase velocity $v_p$ of EAWs. Using the boundary conditions, namely, $\psi^\prime,~d\psi^\prime/d\xi^\prime,~d^2\psi^\prime/d\xi^{\prime2}\rightarrow0$ as $\xi^\prime\rightarrow\pm\infty$, we obtain the following soliton solution.
\begin{equation}
\psi^\prime=\psi^\prime_m\rm{sech}^2\left(\frac{\xi^\prime}{w^\prime}\right),\label{eq-soliton1}
\end{equation}
where $\psi^\prime_m=3u^\prime_0/\cos\alpha$ represents the maximum amplitude and $w^\prime=2\left(\sigma_2l_z^2+\sigma_3l_x^2\right)\sqrt{\cos\alpha/u^\prime_0}$ is the width such that the relation $\psi^\prime_m w^{\prime2}=12$ holds. The solution \eqref{eq-soliton1} can be expressed as \cite{allen1995}
\begin{equation}\label{eq-soliton2}
\psi^\prime=12\eta^2\rm{sech}^2\left[\eta\left\lbrace \left(Z^\prime-4\eta^2T^\prime\right)\cos\alpha+X^\prime\sin\alpha \right\rbrace\right],
\end{equation}
where $\eta=1/w^\prime$.
The soliton energy, i.e., the soliton photon number  is given by
\begin{equation} 
{\cal E}^\prime=\int_{-\infty}^{\infty} |\psi^\prime|^2d\xi^\prime=\frac{4}{3}\psi_m^{\prime2}w^\prime=24\left(u_0\sec\alpha\right)^{3/2}.
\end{equation} \label{eq-sol-energy1}
In terms of the original variables  and in the $\xi$-frame, where $\xi=l_xX+l_zZ-u_0T$, we have the soliton amplitude, width and the energy as 
\begin{equation}\label{eq-sol-energy2}
\begin{split}
&\psi_m\equiv\frac{\psi^\prime_m}{A_1}=\frac{3u_0}{a_1},\\
&w=w^\prime\sqrt{A_2\cos^2\alpha+A_3\sin^2\alpha}=2\sqrt{\frac{a_2}{u_0}},\\
&{\cal E}=\int_{-\infty}^{\infty} |\psi|^2d\xi=\frac{4}{3}\psi_m^{2}w=\frac{24}{a_1^2}\sqrt{a_2u_0^3},
\end{split}
\end{equation}
where $a_1=A_1\cos\alpha$ and $a_2=\left(A_2\cos^2\alpha+A_3\sin^2\alpha\right)\cos\alpha$.
\par 
We note that while the soliton amplitude $\psi_m$ is directly proportional to the soliton speed $u_0$, the width $w$ varies inversely with $u_0$. So, it follows that faster (slower) solitons tend to be taller (shorter) and narrower (wider). Also, from the expression of the soliton energy ${\cal E}$  with its dependency on the soliton amplitude and width, it is evident that electron-acoustic solitons with higher amplitudes (and/or widths) will evolve with higher energies in relativistic degenerate magnetoplasmas. Furthermore considering the electron-acoustic mode as the solitary wave, we note that the soliton solution \eqref{eq-soliton2} gives a propagation speed of $v_p+\varepsilon4\eta^2$, showing an increase in the speed (above the linear phase velocity of EAWs) with decreasing width or increasing amplitude. The latter gives a positive dispersion, i.e., longer-wavelength electron-acoustic solitons (taller and narrower) travel faster than those with shorter wavelengths (shorter and wider). Thus, it follows that when the conditions for the formation of solitons are fulfilled (i.e., the steepening associated with the wave nonlinearity is nicely balanced with the dispersion broadening the pulse), solitons travel faster than the phase speed of the corresponding linear EAW. 
\par  
\begin{figure*} 
	\centering
	\includegraphics[width=6.5in,height=3.2in]{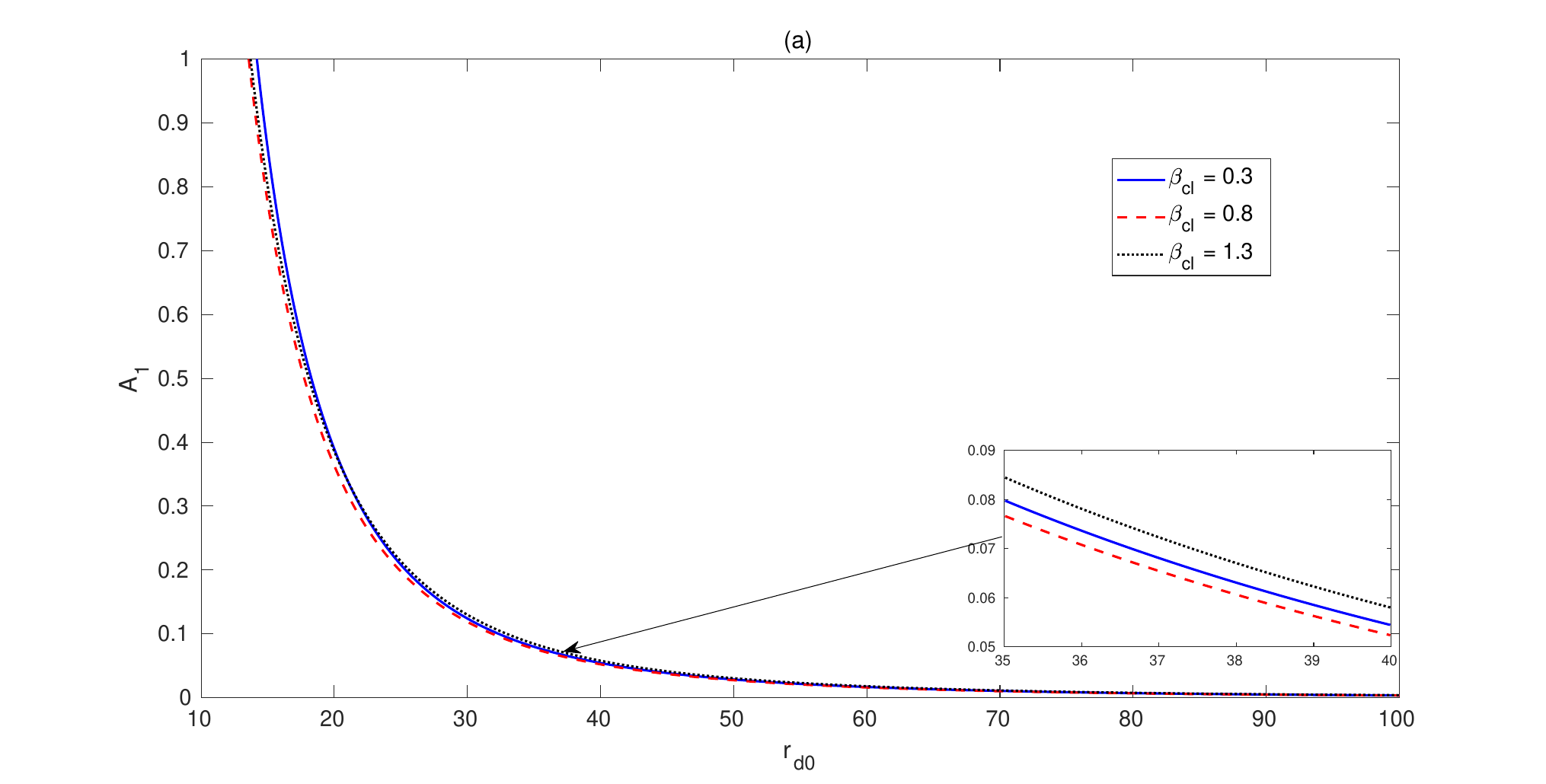}\\
	\includegraphics[width=6.5in,height=3.2in]{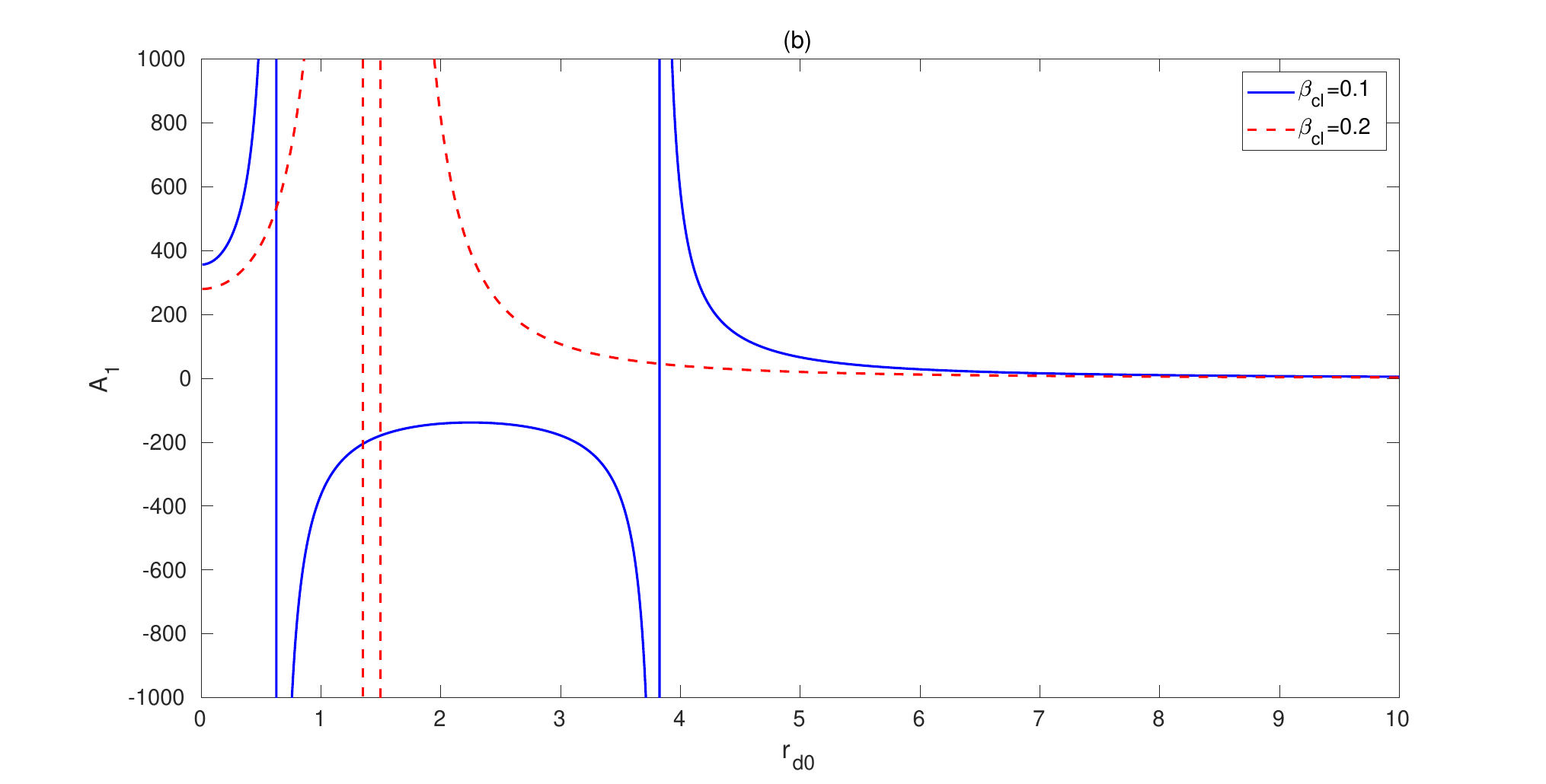}
	\caption{ The nonlinear coefficient $A_1$ of the ZK equation is plotted against $r_{d0}$ for different values of $\beta_{\rm{cl}}$ as in the legends to show the regimes for $A_1>0$ and $A_1<0$ for which electron-acoustic solitons with positive or negative potentials can exit. Subplot (a) shows that $A_1$ can be positive in the entire domain of $r_{d0}$ but with relatively higher values of $\beta_{\rm{cl}}$. However, $A_1$ can be negative at some lower values of $\beta_{\rm{cl}}$ [Subplot (b)]. }  \label{fig2-A1}
\end{figure*}
From Eq. \eqref{eq-sol-energy2}, we note that the amplitude of electron-acoustic solitons typically depends on the nonlinear coefficient $A_1$, which again depends parametrically on $\beta_{\rm{cl}}$ and $r_{d0}$. So, the solitons are rarefactive ($\psi_m<0$) or compressive ($\psi_m>0$) according to when $A_1<0$ or $A_1>0$. However, when $A_1$ is close to zero or vanishes in the parameter space, the ZK equation \eqref{eq-ZK} fails to describe the evolution of electron-acoustic solitons. In that case, we require to derive a modified ZK equation by considering higher-order perturbations of the dependent physical variables or slower space and time scales in the new coordinate frame of reference. 
\par
\begin{figure*}
	\centering
	\includegraphics[width=7in,height=2.5in]{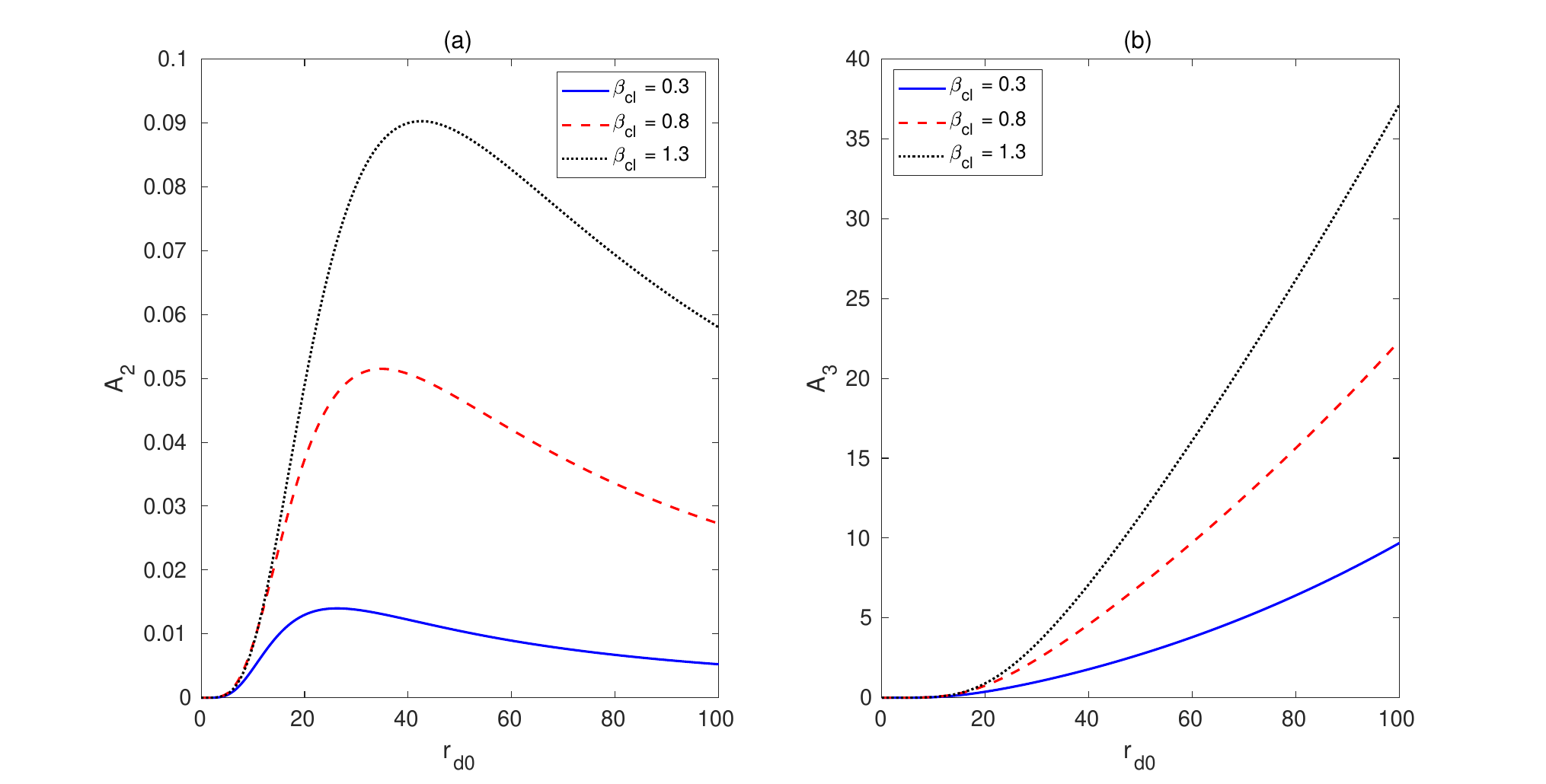}
	\caption{The logitudinal [$A_2$, subplot (a)] and transeverse [$A_3$, subplot (b)] dispersion coefficients of the ZK equation are plotted against $r_{d0}$ for different values of $\beta_{\rm{cl}}$ as in the legends. The soliton profile and its energy are greatly influenced by the characteristics of these coefficients.  }  \label{fig3-A2-A3}
\end{figure*}

 
We numerically investigate the nature of the coefficients $A_1$, $A_2$, and $A_3$ of the ZK equation against the degeneracy parameter $r_{d0}$ for different values of the relativity parameter $\beta_{\rm{cl}}$. The results are displayed in Figs. \ref{fig2-A1} and \ref{fig3-A2-A3}. We find that when $\beta_{\rm{cl}}\gtrsim0.3$ [see Fig. \ref{fig2-A1} (a)], $A_1$ is positive in a wide range of values of $r_{d0}$. However, it approaches to zero for $r_{d0}\gtrsim50$, i.e., in the strong relativistic degeneracy regime. In the latter, the ZK equation \eqref{eq-ZK} may not be valid for the description of electron-acoustic solitons. From the subplot (a) of Fig. \ref{fig2-A1}, we also note that the values of $A_1$ increase with increasing values of $\beta_{\rm{cl}}$, i.e., the nonlinearity associated with the wave steepening enhances as the thermal energy of classical electrons tends to exceed the electron rest mass energy. On the other hand, for $\beta_{\rm{cl}}<0.3$ [see Fig. \ref{fig2-A1} (b)], $A_1$ can be negative in a small interval of $r_{d0}$ and as the values of $\beta_{\rm{cl}}$ are reduced  from $\beta_{\rm{cl}}=0.3$, such interval expands with higher values of $r_{d0}$. For example, for $\beta_{\rm{cl}}=0.2$, $A_1<0$  in  $1.36\lesssim r_{d0}\lesssim1.48$ and for $\beta_{\rm{cl}}=0.1$, $A_1<0$ in $0.5<r_{d0}<4$. Thus, it follows that if the temperature inside the core of white dwarfs is typically \cite{saumon2022,kryvdyk2007,kryvdyk1999,misra2012}  $\sim10^8$ K and the number density of degenerate electrons varies in $7\times10^{28}\lesssim n_{d0}\lesssim4\times10^{31}~\rm{cm}^{-3}$, the electron-acoustic solitons with negative potential are more likely to form. 
It is to be mentioned that the results shown in Figs. \ref{fig2-A1} and \ref{fig3-A2-A3} are for strongly magnetized plasmas with $\Omega_e\gtrsim50$ at which the coefficients of the ZK equation can assume low to  moderate values. However, in the regimes of weak magnetic fields with $\Omega_e\lesssim1$, the dispersion coefficient $A_3$ becomes significantly high (as is evident from its expression) compared to $A_1$ and $A_2$, for which the relevant dynamics may not be so important, especially in the environments of compact astrophysical objects. 
Figure \ref{fig3-A2-A3} shows that the dispersion associated with the transverse (to the direction of propagation) perturbation always dominates over that due to the longitudinal perturbation. Also, the parameter $\beta_{\rm{cl}}$ significantly influences on their magnitudes, i.e., both the values of $A_2$ and $A_3$ increase with the increasing values of 
$\beta_{\rm{cl}}$. Furthermore, both $A_2$ and $A_3$ are positive in a wide range of values of $r_{d0}$ that correspond from weakly to ultra-relativistic degeneracy regimes. Also, while the values of $A_3$ increase  with increasing values of $r_{d0}$, those of $A_2$ initially increase in a sub-interval of $r_{d0}$ and then decrease in the rest of the interval.   From Figs. \ref{fig2-A1} and \ref{fig3-A2-A3}, it may be predicted that the soliton amplitude (which is inversely proportional to $A_1$) increases with increasing values of $r_{d0}$ (with a fixed value of $\beta_{\rm{cl}}$) but it  decreases (increases) with increasing value of $\beta_{\rm{cl}}>1~(<1)$ (with a fixed value of $r_{d0}$).  On the other hand, since the soliton width is directly proportional to both $A_2$ and $A_3$ and the magnitudes of $A_3$ are much higher than $A_2$,  they increase with increasing values of both $r_{d0}$ and $\beta_{\rm{cl}}$. 
\par
\begin{figure*}  
	\centering
	\includegraphics[width=7in,height=2.5in]{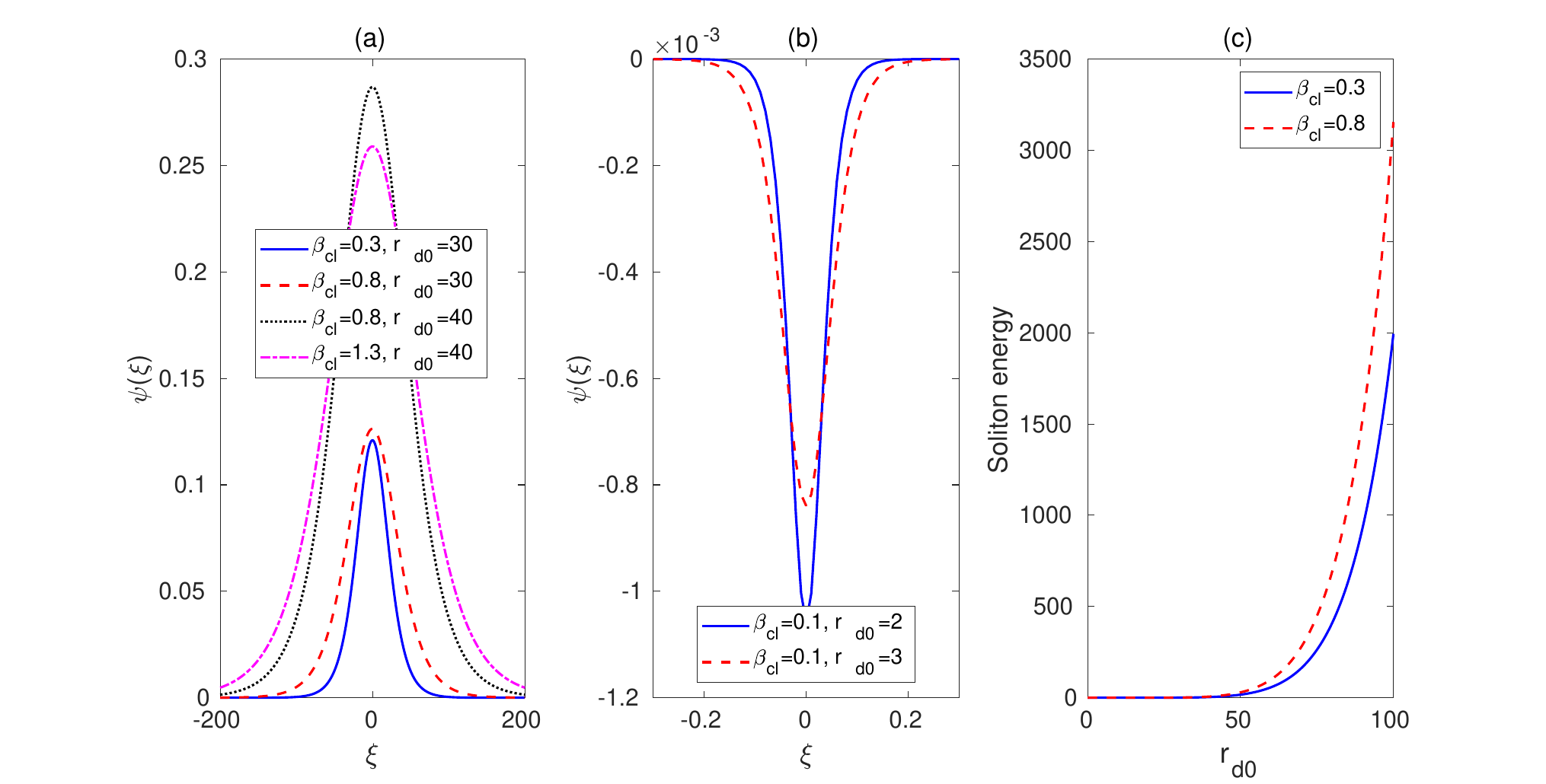}
	\caption{The profiles of the ZK solitons [Subplots (a) and (b) correspond to compressive and rarefactive solitons respectively] and soliton energy [subplot (c)] are shown with the variations of $r_{d0}$ and  $\beta_{\rm{cl}}$ as in the legends.     } 
	\label{fig4-sol-energy}
\end{figure*}

 
Having analyzed the profiles of the coefficients of the ZK equation and the parameter regimes for the existence of compressive and rarefactive solitons, we plot the characteristics of electron-acoustic solitons and the soliton energy for different values 
of $r_{d0}$ and $\beta_{\rm{cl}}$ as shown in Fig. \ref{fig4-sol-energy}. We find that the relativity and the degeneracy parameters significantly affect both the amplitude and width of solitons, thereby changing their polarity and energy. From Fig. \ref{fig4-sol-energy}, it is confirmed that in the case of solitons with positive polarity (compressive type), both the amplitude and width of solitons increase with increasing values of both $r_{d0}$ and $\beta_{\rm{cl}}<1$. However, when $\beta_{\rm{cl}}$ increases from a value smaller than unity to values larger than unity  [see the dotted and dash-dotted lines in subplot (a)], the width keeps increasing  but the amplitude decreases.   The increments of both the amplitude and width are, however, noticeable with an enhancement of the degeneracy parameter $r_{d0}$ [see the dashed and dotted lines in subplot (a)]. On the other hand, for solitons with negative polarity [subplot (b)], it is seen that the effect of $r_{d0}$ is to enhance the width but to decrease their amplitudes (in magnitudes). From subplot (c), it is also evident that the soliton energy increases with increasing values of both $r_{d0}$ and  $\beta_{\rm{cl}}<1$. However, the increment is insignificant for $\beta_{\rm{cl}}>1$. Thus, it may be concluded that high-energy electron-acoustic solitons can propagate in ultra-relativistic degenerate plasmas with a mixture of low-density classical electrons having thermal energy more or less than the rest mass energy where the temperature of classical electrons can vary in the range \cite{saumon2022,valyavin2011,kryvdyk2007,kryvdyk1999,misra2012}  $10^7\lesssim T_{\rm{cl}}\lesssim10^9$ K and the number density of degenerate electrons in   $7\times10^{28}\lesssim n_{d0}\lesssim10^{34}~\rm{cm}^{-3}$ or more. Also, the solitons are more likely to appear of the compressive type in higher density regimes ($n_{d0}\gtrsim10^{31}~\rm{cm}^{-3}$) with higher temperature ($T_{\rm{cl}}\gtrsim10^8$ K). However, in relatively low temperature regimes \cite{misra2012} ($\lesssim10^8$ K) with density  $7\times10^{28}\lesssim n_{d0}\lesssim4\times10^{31}~\rm{cm}^{-3}$, the rarefative type solitons can appear.
\section{Modified ZK equation and its soliton solution}\label{sec-mZK}
From the analysis in Sec. \ref{sec-ZK}, we have observed that the nonlinear coefficient $A_1$ tends to vanish (but not exactly zero) for $r_{d0}\gtrsim50$, i.e., in strong relativistic degenerate plasmas. In this regime, the ZK equation \eqref{eq-ZK} fails to describe the evolution of electron-acoustic solitons. Thus, it is necessary to consider higher order perturbations together with slower space and time scales in the stretched coordinates in comparison with those used for the ZK equation. So, we assume that $A_1\approx \sigma_1\varepsilon$, where $\sigma_1=\rm{sign}(A_1)=\pm1$ according to when $A_1\gtrless0$. Some modifications of the stretched coordinates and the expansion for the transverse components of the fluid velocity are also required, which give
 \begin{eqnarray}
 \begin{split}
&X=\varepsilon x,~Z=\varepsilon\gamma_v\left(z-v_pt\right),\\
&T=\varepsilon^3\gamma_v\left(t-v_pz\right),\label{eq-str-new}
\end{split}
\end{eqnarray}
 \begin{equation}
 \begin{split}
&u_x=\varepsilon^2{u}_x^{(1)}+\varepsilon^3{u}_x^{(2)}\cdots, \\
&u_y=\varepsilon^2{u}_y^{(1)}+\varepsilon^3{u}_y^{(2)}\cdots.\label{eq-expan-new}
\end{split}
\end{equation}
Substitution of Eqs. \eqref{eq-str-new} and \eqref{eq-expan-new} in the basic equations \eqref{eq-cont1}-\eqref{eq-poiss1} gives the same results for the first-order perturbed quantities. However, for the  second-order perturbations, we have
\begin{widetext}
\begin{equation}
n_d^{\left(2\right)}=\frac{3}{2\left(r_{d0}^2-3R_0^3v_p^2\right)^3}\Biggl(r_{d0}^2\left(R_2^2+3R_0^2\left(1+5R_0\right)v_p^2\right)
+9R_0^3v_p^2\left(1+R_0^2\left(-3+v_p^2\right)\right)\Biggr)\left(\phi^{\left(1\right)}\right)^2+\frac{3R_0\phi^{\left(2\right)}}{r_{d0}^2-3R_0^3v_p^2},
\end{equation}
\begin{equation}
n_{cl}^{\left(2\right)}=\frac{\left(G_{cl}v_p^2\left(-2+v_p^2\right)+\beta_{cl}\left(1+v_p^2\right)\right)}{2\left(\beta_{cl}-G_{cl}v_p^2\right)^3}\left(\phi^{\left(1\right)}\right)^2
+\frac{\phi^{\left(2\right)}}{\beta_{cl}-G_{cl}v_p^2}, 
\end{equation}
\begin{equation}
u_{dz}^{(2)}=\frac{3v_p}{2\left(r_{d0}^2-3R_0^3v_p^2\right)^3}\Biggl[\Bigl(r_{d0}^2R_2^2-9R_0^5v_p^2+3R_0^3\left(3+5r_{d0}^2\right)v_p^2+6R_0^2r_{d0}^2\left(-1+v_p^2\right)\Bigr)\left(\phi^{\left(1\right)}\right)^2+2R_0\left(r_{d0}^2-3R_0^3v_p^2\right)^2\phi^{\left(2\right)}\Biggr], 
\end{equation}
\begin{equation}
u_{clz}^{\left(2\right)}=\frac{v_p}{2\left(\beta_{cl}-G_{cl}v_p^2\right)^3}\Biggl[\beta_{cl}\left(-1+2v_p^2\right)\left(\phi^{\left(1\right)}\right)^2+2\left(\beta_{cl}-G_{cl}v_p^2\right)^2\phi^{\left(2\right)}\Biggr].\label{eqd41}
\end{equation}
\end{widetext}
The Poisson equation \eqref{eq-poiss1} gives
\begin{widetext}
\begin{equation}
\alpha_1\phi^{\left(2\right)}=-\frac{1}{2\delta}\left[\frac{\left(\beta_{cl}+2\left(-G_{cl}+\beta_{cl}\right)v_p^2\right)\delta}{\left(\beta_{cl}-G_{cl}v_p^2\right)^3}
+\frac{3r_{d0}^2R_2^2+9R_0^2\left(-9R_0^3+2r_{d0}^2+R_0\left(3+5r_{d0}^2\right)\right)v_p^2}{\left(r_{d0}^2-3R_0^3v_p^2\right)^3}\right]\left(\phi^{\left(1\right)}\right)^2, \label{eq-phi2}
\end{equation}
\end{widetext}
where $\alpha_1$, the coefficient of $\phi^{\left(2\right)}$, is given by
\begin{equation}
\alpha_1=\frac{1}{G_{\rm{cl}}v_p^2-\beta_{\rm{cl}}}-\frac{3R_0}{\delta\left(r^2_{d0}-3R_0^3v_p^2\right)}.
\end{equation}
 Equation \eqref{eq-phi2} can be rewritten in the form $\alpha_1\phi^{\left(2\right)}=Q\ \left(\phi^{\left(1\right)}\right)^2$. Since $\alpha_1$ can be made identically zero by the linear dispersion relation \eqref{eq-phase} and $\phi^{\left(1\right)}\neq0$, it follows that $Q$ should be at least of the  order of $\varepsilon$, and the term proportional to $Q$ will appear in the third-order correction terms of the Poisson equation. So, considering the third-order perturbed quantities, we finally obtain, after a few steps, the following modified ZK (mZK) equation.  
\begin{equation}
\begin{split}
\frac{\partial\psi}{\partial T}+&\sigma_1\psi\frac{\partial\psi}{\partial Z}+A_4\psi^2\frac{\partial\psi}{\partial Z}\\+&A_2\frac{\partial^3\psi}{\partial Z^3}+A_3\frac{\partial^3\psi}{\partial Z\partial X^2}=0,\label{eq-mZK}
\end{split}
\end{equation}
where $\psi=\phi^{(1)}$ and the new nonlinear coefficient $A_4$ is given by
\begin{widetext}
\begin{eqnarray}
A_4=&&\frac{\gamma_v^2}{R_0\left(p_{cl0}-G_{cl}v_p^2\right)D}\left[\frac{3R_0^2v_p^2\left[r_{d0}^2-2R_0\left(3+2r_{d0}^2\right)\right]}{\left(r_{d0}^2-3R_0^3 v_p^2\right)^2}
-\frac{R_0\left[\beta_{cl}+2\left(\beta_{cl}-G_{cl}\right)v_p^2\right]}{\left(\beta_{cl}-G_{cl}v_p^2\right)^2}\right.\nonumber \\
&&\left.+\frac{r_{d0}^2\left(R_2^2+3R_0^2v_p^2\right)}{\left(r_{d0}^2-3R_0^3v_p^2\right)^2}\right].\label{eq-A4}
\end{eqnarray}
\end{widetext}
The other coefficients $A_1$, $A_2$, and $A_3$ are the same as given before.
\par 
By comparing Eqs. \eqref{eq-ZK} and \eqref{eq-mZK}, we find that an additional                                                                         nonlinear term proportional to $A_4$ appears in Eq. \eqref{eq-mZK} due to the smallness of $A_1\sim{\cal O}(\varepsilon)$ and consideration of the higher-order perturbations of the electrostatic potential. Equation \eqref{eq-mZK} is valid for parameter values at which $A_1$ is close to zero but not exactly zero, i.e., in the regimes of high relativistic degeneracy with the degeneracy parameter satisfying $r_{d0}\gtrsim50$.   
\par 
To obtain a plane soliton solution of the mZK equation \eqref{eq-mZK}, we apply the transformation
\begin{equation}
\psi\rightarrow \sqrt{A_2}\psi,~T\rightarrow T/A_2^2,~X\rightarrow X/A_2\sqrt{A_3},~Z\rightarrow Z/A_2^{3/2},
\end{equation}
where as shown before, $A_2,~A_3>0$. Thus, Eq. \eqref{eq-mZK} reduces to [after rewriting $(X,Z,T)$ as $(x,z,t)$]
\begin{equation}
\frac{\partial\psi}{\partial t}+\sigma_1\psi\frac{\partial\psi}{\partial z}+B\psi^2\frac{\partial\psi}{\partial z}+\frac{\partial}{\partial z}\left(\frac{\partial^2}{\partial z^{2}}+\frac{\partial^2}{\partial x^{2}}\right)\psi=0,\label{eq-mZK2}
\end{equation}
where $B=A_4/\sqrt{A_2}$. A traveling solitary wave solution of Eq. \eqref{eq-mZK2} in a frame moving with a constant speed $u_0$ can be obtained as \cite{hongsit2008}
\begin{equation}
\psi(x,z,t)=\frac{6\sigma_1u_0}{1+(1/\beta)\cosh\eta\left(l_xx+l_zz-u_0t\right)}, \label{eq-sol1-mzk}
\end{equation}
where $\eta=\sqrt{u_0/l_z}$, $\beta=1/\sqrt{1+6u_0B}$, $l_x=\sin\alpha$, and $l_z=\cos\alpha$. Since $A_4$ is found to be negative in wide ranges of values of $r_{d0}$ and $\beta_{\rm{cl}}$, for real values of $\beta$ one must have $1+6u_0B>0$.   The soliton energy is obtained as
 \begin{equation}\label{eq-energy-mzk}
 {\cal E}=\int_{-\infty}^{\infty} |\psi|^2d\xi=\frac{4}{\eta}u_0\beta^2,
 \end{equation}
 where $\xi=l_xx+l_zz-u_0t$.
 \par
\begin{figure*}  
	\centering
	\includegraphics[width=6in,height=2.5in]{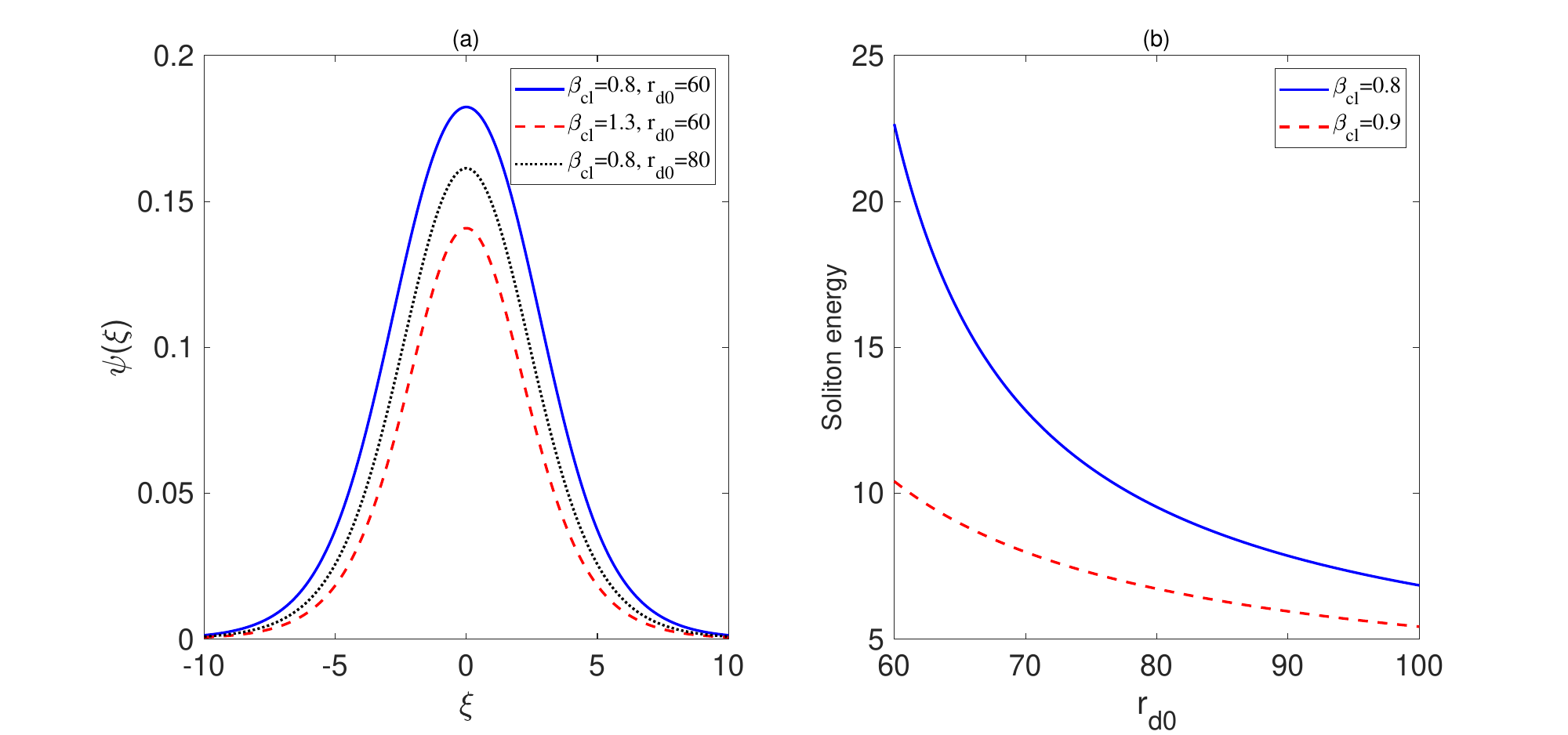}
	\caption{The profiles of the mZK soliton [Eq. \eqref{eq-sol1-mzk}, subplot (a)] and the soliton energy [Eq. \eqref{eq-energy-mzk}, subplot (b)] are shown for different values of $r_{d0}$ and $\beta_{{\rm{cl}}}$ as in the legends. In subplot (a), the absolute values of $\psi$ are considered.} 
	\label{fig5-sol-energym}
\end{figure*}

 
 Like ZK solitons, the propagation speed of mZK solitons can also be shown to be larger than the phase speed of EAWs, i.e., mZK solitons can have postive dispersion. The profiles of the mZK soliton and the soliton energy are shown in Fig. \ref{fig5-sol-energym}. It is found that although the effect of $\beta_{\rm{cl}}$ on the profile of the mZK soliton is similar to the ZK soliton,  both the amplitude and width of the mZK soliton decrease (in contrast to the ZK soliton)  with increasing values of the degeneracy parameter $r_{d0}$ [See subplot (a)]. Also, in contrast to the ZK soliton,  the energy of the mZK soliton tend to decrease with increasing values of $r_{d0}$ and the decrement is significant with a small enhancement of $\beta_{\rm{cl}}$ [see subplot (b)].  Thus, electron-acoustic solitons, which may appear as compressive or rarefactive type, can propagate in degenerate dense magnetoplasmas. However, they tend to lose energy in the regimes of ultra-relativistic degeneracy, such as those in the environments of white dwarfs in which the degenerate core contains the bulk of the mass, surrounded by a thin non-degenerate envelope.   
\section{Stability of ZK and mZK solitons}\label{sec-stability}
We note that in contrast to the KdV equation, the ZK and the mZK equations are not integrable. So, it may not be possible to fully describe  the evolution process of plane ZK and mZK solitons analytically. Instead, one can investigate the transition processes of these solitons in step by step of which the first one is the linear stability analysis. We perform this analysis for both the ZK and mZK solitons following the works of Allen and Rowlands \cite{allen1995}  and Hongsit \textit{et al.} \cite{hongsit2008}   in the limit of small wave numbers $(k\ll1)$. The latter correspond to low-frequency long-wavelength electron-acoustic perturbations.
\subsection*{Stability of ZK soliton}
We rewrite the ZK equation \eqref{eq-ZK-nond1} after dropping the primes and rewriting $(X,Z,T)$ as $(x,z,t)$ for brevity, as 
\begin{equation}
\frac{\partial\psi}{\partial t}+\psi\frac{\partial\psi}{\partial z}+\frac{\partial}{\partial z}\left(\frac{\partial^2}{\partial z^{2}}+\frac{\partial^2}{\partial x^{2}}\right)\psi=0.\label{eq-ZK-nond2}
\end{equation}
For electron-acoustic solitary waves propagating at angle $\alpha$ with the $Z$-axis or the direction of the external magnetic field, we apply the transformation
\begin{equation}
 x^\prime=\eta x,~z^\prime=\eta\left(z-4\eta^2 t\right),~t^\prime=\eta^3t,~\psi^\prime=\frac{\psi}{\eta^2}
\end{equation}
to Eq. \eqref{eq-ZK-nond2} and then use the orthogonal transformation
\begin{equation}
\left(x^{\prime\prime},z^{\prime\prime}\right)=\left(x^\prime\cos\alpha-z^\prime\sin\alpha,x^\prime\sin\alpha+z^\prime\cos\alpha\right)  
\end{equation}
to rotate the axes. The resulting equation after dropping the primes is
\begin{equation}\label{eq-zk-nond0}
\begin{split}
\frac{\partial\psi}{\partial t}+&\left(\cos\alpha\frac{\partial}{\partial z}-\sin\alpha\frac{\partial}{\partial x}\right)\\
&\times \left(\frac{1}{2}\psi^2-4\psi+\frac{\partial^2 \psi }{\partial x^{2}}+\frac{\partial^2 \psi }{\partial z^{2}}\right)=0.
\end{split}
\end{equation} 
Clearly, Eq. \eqref{eq-zk-nond0} has a plane soliton solution traveling along the $z$-axis in the following form.
\begin{equation}
\psi_0(z)=12~\rm{sech}^2z. \label{eq-sol-zk3}
\end{equation}
Next, we perturb the solution $\psi(x,z,t)$ of Eq. \eqref{eq-zk-nond0} as $\psi(x,z,t)=\psi_0(z)+\psi_1(x,z,t)$, where we consider the perturbation, $\psi_1$ in the form of a plane wave. Thus, we write
\begin{eqnarray}
\psi(x,z,t)=\psi_0(z)+\varepsilon \tilde{\psi}(z)\exp{\left(ikx-i\omega t\right)}, \label{eq-sol-pert}
\end{eqnarray}   
where $\varepsilon~(\ll k^2)$ denotes a small scale of perturbation. Also, $k$ $(\omega)$  is the wave number (frequency)  of perturbation and  $\omega$ is, in general, complex such that $\Im\omega~(>0)$ represents the growth rate of instability and  $\Re\omega$ contributes to the phase of the plane wave. Thus, stationary ZK solitons under the plane wave perturbation are said to be stable when $\Im\omega=0$ and unstable when $\Im\omega~(>0)$. Using the multiple-scale perturbation technique as in Ref. \cite{allen1995}, i.e., considering $z_1=kz,~z_2=k^2z$, etc. and assuming $k\ll1$, we expand $\tilde{\psi}(z)$ and $\omega$ as 
\begin{equation}
\begin{split}
&\tilde{\psi}(z)=\tilde{\psi}_0+k\tilde{\psi}_1+k^2\tilde{\psi}_2+..., \label{eq-expansion1}
\omega=k\omega_1+k^2\omega_2+... 
\end{split}
\end{equation}
In what follows, we substitute Eq. \eqref{eq-sol-pert} into Eq. \eqref{eq-zk-nond0} and linearize with respect to $\varepsilon$. Next, substituting the expansion \eqref{eq-expansion1} into the resulting linearized equations and collecting ascending powers of $k$, we obtain, after few steps, the following expressions for the first and second-order growth rates of instability of the perturbation $\psi_1$.
\begin{equation}
\omega_1=\frac{8}{3}\left[i\left(\frac{8}{5}{\cos}^2{\alpha}-1\right)^{1/2}-\sin{\alpha}\right],\label{eq-Gam1}
\end{equation}
\begin{equation}
\omega_2=\frac{4}{9}i\left(1-\frac{8}{5}{\cos}^2{\alpha}\right)\sec{\alpha}-\frac{4}{45}\frac{\left(5+4{cos}^2{\alpha}\right)   \tan{\alpha}}{\left(\frac{8}{5}{\cos}^2{\alpha}-1\right)^{1/2}}.\label{eq-Gam2}
\end{equation}
Since $-1<\cos\alpha<1$, one can easily find from Eq. \eqref{eq-Gam1}  that the first-order soliton stability is ensured when $(8/5)\cos^2\alpha-1<0$ or $\alpha>37.8^\circ$ and the first-order instability occurs when $\Im\omega_1>0$, i.e., when $\alpha<37.8^\circ$. However, from Eq. \eqref{eq-Gam2} it is interesting to note that the soliton can be unstable in the second-order correction even if it is stable for the first-order perturbations.  So,  when $\alpha>37.8^\circ$, even though the first-order stability is ensured, we can still have the second-order instability with $\Im\omega_2>0$.    The growth rate of instability in two different cases can be summarized as follows:\\
(i) When $\alpha<\alpha_{\rm{cr}}\cong37.8^\circ$, the growth rate of instability  is given by
\begin{eqnarray}
\Gamma_1=k\Im{\omega_1}+{\cal O}(k^2)\cong\frac{8}{3}k\left(\frac{8}{5}{\cos}^2{\alpha}-1\right)^{1/2},\label{eqd44}
\end{eqnarray}
(ii) When $\alpha>\alpha_{\rm{cr}}\cong37.8^\circ$,   the growth rate is given by
\begin{equation}
\begin{split}
\Gamma_2=&k^2\Im{\omega_2}+{\cal O}(k^3)\\
&\cong\frac{4}{9}k^2\left(1-\frac{8}{5}{\cos}^2{\alpha}\right)\sec{\alpha}.\label{eqd45}
\end{split}
\end{equation}
Reverting back to the stability of solitons corresponding to the ZK equation \eqref{eq-ZK}, i.e., making use of the transformation 
\begin{equation}
t\rightarrow \frac{8T}{\sqrt{A_2}}\left(\frac{\cos\alpha}{u_0}\right)^{3/2},
\end{equation}
we obtain the instability growth rates as
\begin{eqnarray}
\Gamma_1\cong  \frac{1}{3}\frac{k}{\sqrt{A_2}}\left(u_0\sec\alpha\right)^{3/2}\left(\frac{8}{5}{\cos}^2{\alpha}-1\right)^{1/2},\label{eq-growth-ZK1}
\end{eqnarray}
\begin{eqnarray}
\Gamma_2\cong \frac{1}{18}\frac{k^2}{\sqrt{A_2}}\left(u_0\sec^{5/3}\alpha\right)^{3/2}\left(1-\frac{8}{5}{\cos}^2{\alpha}\right),\label{eq-growth-ZK2}
\end{eqnarray}
where $k$ has the role of a scaling parameter of smallness of perturbation.
\begin{figure*}
	\centering
	\includegraphics[width=6in,height=2.5in]{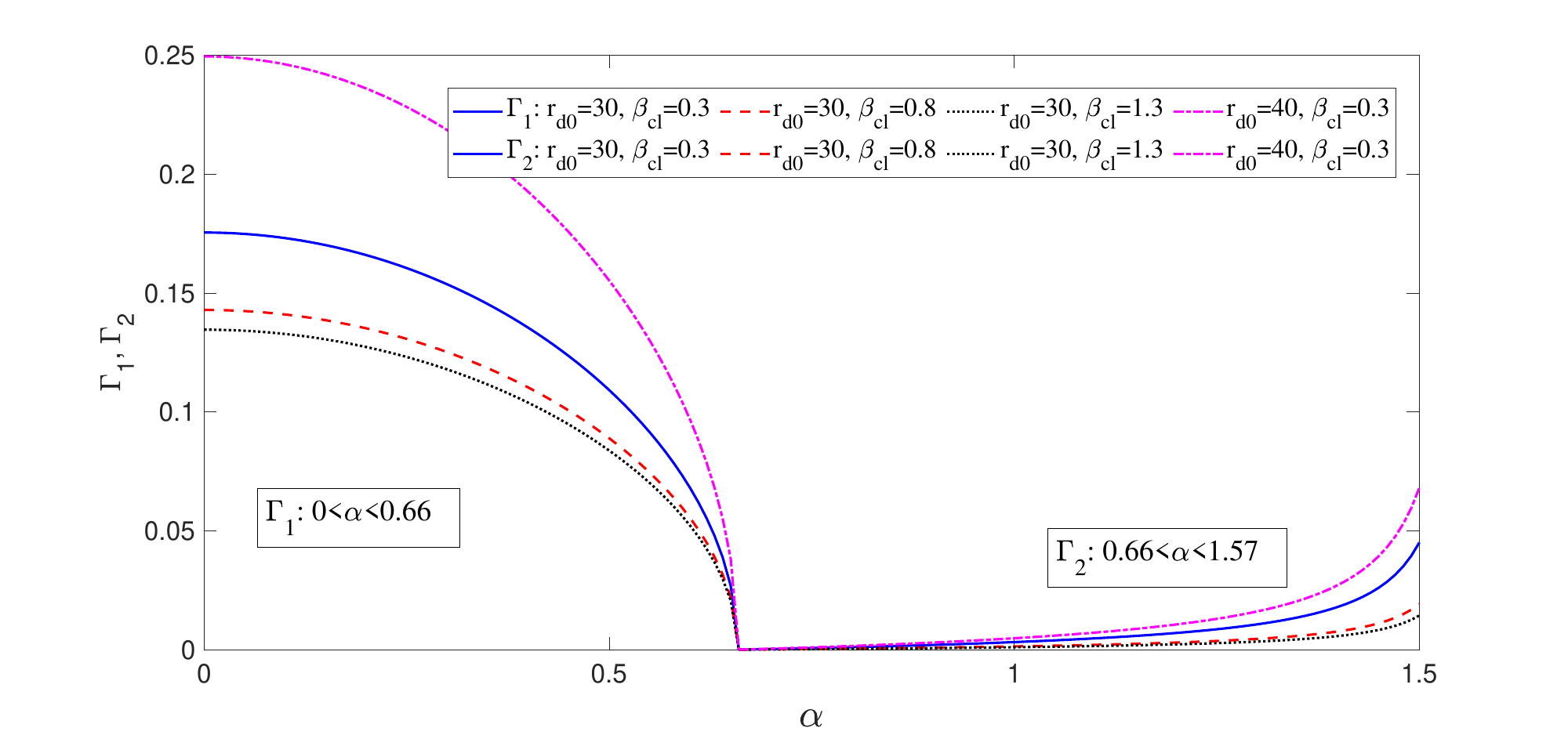}
	\caption{The first and second-order growth rates of instability of the ZK soliton [Eqs. \eqref{eq-growth-ZK1} and \eqref{eq-growth-ZK2}] are shown for different values of the relativistic degeneracy parameter $r_{d0}$ and the relativity parameter $\beta_{\rm{cl}}$ as in the legends.   }  \label{fig6-growthrate-zk}
\end{figure*}
The profiles of the first and second-order instability growth rates are shown in Fig. \ref{fig6-growthrate-zk}. It is seen that although the first-order growth rate tends to decrease with the angle of propagation of solitons ($\alpha$) having cut-off at a value close to  $\alpha=37.8^\circ$, the second-order growth rate, in contrast, increases with $\alpha$. In both the cases, the instability growth rate is significantly reduced with an increase of the relativity parameter $\beta_{\rm{cl}}$ (see the solid, dashed, and dash-dotted lines). However, as one enters the regime of strong relativistic degeneracy, the same increases with $\beta_{\rm{cl}}$.  It is to be mentioned that the critical angle $\alpha_c=37.8^\circ$ for the stability of solitons is independent of the plasma parameters and applicable to ZK solitons in other plasma environments (See, e.g., Ref. \cite{shatashvili2020}). Such critical angle can be matched with that obtained by Das and Verheest \cite{das1989} for ZK solitons using the consistency condition reported by Allen and Rowlands \cite{allen1995} in their work. Furthermore, the characteristics of the growth rates of instability associated with the first- and second-order corrections of perturbations are in agreement with those obtained for ZK solitons in classical nonthermal electron-positron-ion plasmas \cite{shatashvili2020}.  
\subsection*{Stability of mZK soliton}                                                                      
\par
Before proceeding to the stability analysis of plane solitary pulses of the mZK equation, we further simplify the solution \eqref{eq-sol1-mzk} by applying the following transformation to Eq. \eqref{eq-mZK2}.
\begin{equation}
\begin{split}
 &x^\prime=\eta x,~z^\prime=\eta\left(z-4\eta^2 t\right),~t^\prime=\eta^3t,\\
 &\psi^\prime=\frac{\psi}{\eta^2},~B^\prime= \eta^2B,
 \end{split}
\end{equation}
 and then use the orthogonal transformation
\begin{equation}
\left(x^{\prime\prime},z^{\prime\prime}\right)=\left(x^\prime\cos\alpha-z^\prime\sin\alpha,x^\prime\sin\alpha+z^\prime\cos\alpha\right)  
\end{equation}
to rotate the axes. The resulting equation after dropping the primes is
\begin{equation} \label{eq-mZK3}
\begin{split}
&\frac{\partial\psi}{\partial t}+\left(\cos\alpha\frac{\partial}{\partial z}-\sin\alpha\frac{\partial}{\partial x}\right)   \\
&\times\left(\frac{s}{2}\psi^2-4\psi+\frac{B}{3}\psi^3+\frac{\partial^2 \psi }{\partial x^{2}}+\frac{\partial^2 \psi }{\partial z^{2}}\right)=0. 
\end{split}
\end{equation} 
Thus, the solitary solution \eqref{eq-sol1-mzk} reduces to
\begin{equation}
\psi_0(z)=\frac{6su_0}{1+(1/\beta)\cosh{z}}. \label{eq-sol2-mzk}
\end{equation}
\par
\begin{figure*}
	\centering
	\includegraphics[width=6in,height=2.5in]{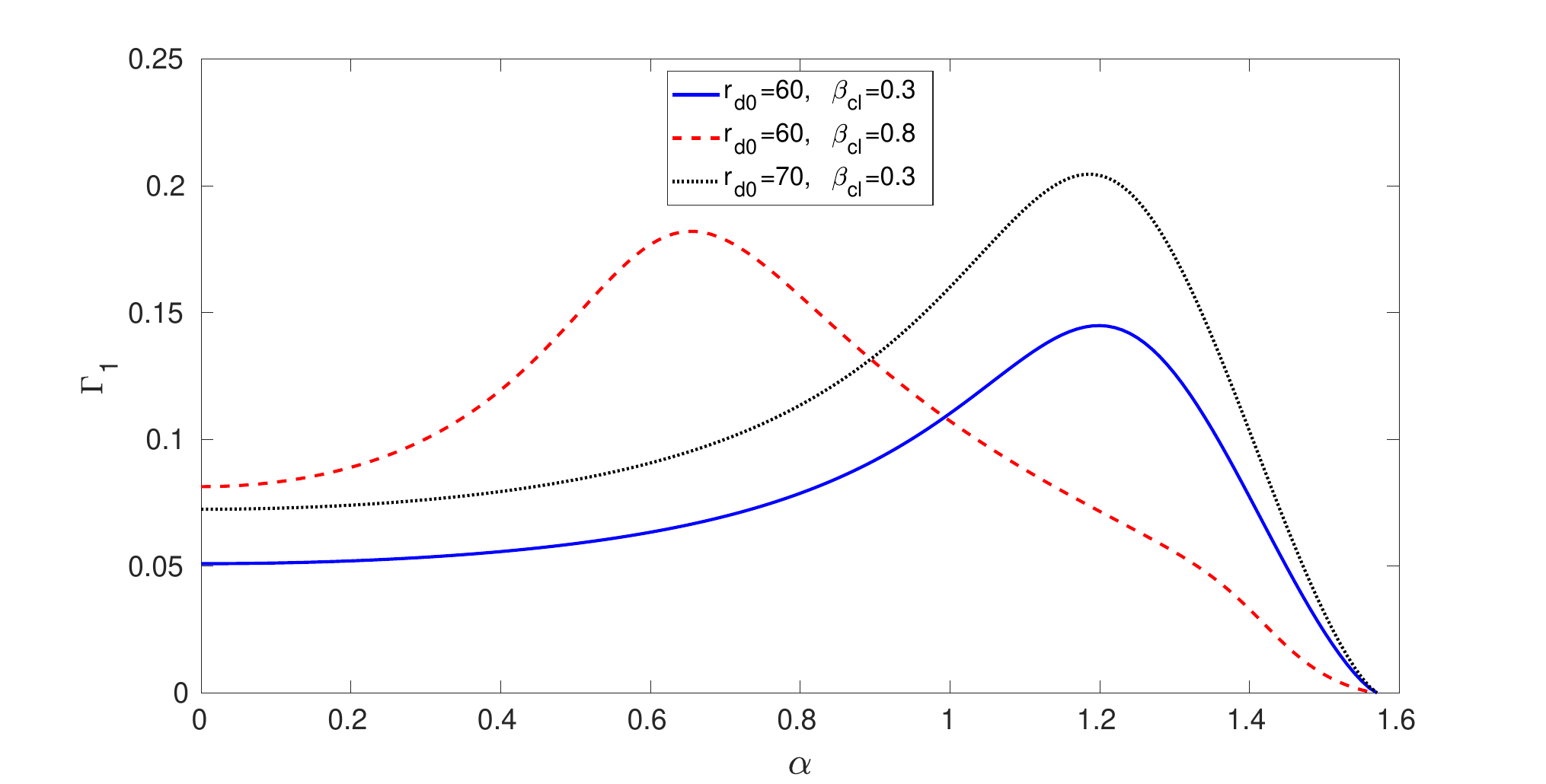}
	\caption{The growth rate of instability of the mZK soliton [Eq. \eqref{eq-growth-mZK}] is shown for different values of the relativistic degeneracy parameter $r_{d0}$ and the relativity parameter $\beta_{\rm{cl}}$ as in the legends.}  \label{fig7-growth-mZK}
\end{figure*}

  
Proceeding in the same way as for the ZK equation, i.e., assuming 
\begin{eqnarray}
\psi(x,z,t)=\psi_0(z)+\varepsilon \tilde{\psi}(z)\exp{\left(ikx-i\omega t\right)},  
\end{eqnarray} 
we obtain from Eq. \eqref{eq-mZK3}, after replacing $B$ by $B/\eta^2$ in the definition of $\beta$ and making use of the transformations $t\rightarrow \eta^3 T/A_2^2$  (for which the normalizing factors for $\omega$ is $\eta^3/A_2^2$), the following expression for the first-order growth rate of instability \cite{hongsit2008}.
\begin{equation} \label{eq-growth-mZK}
\Gamma_1\approx \frac{A_2\sqrt{A_3}}{\sqrt{u_0\sec\alpha}}k\omega_1,
\end{equation}
 where
 \begin{widetext}
\begin{eqnarray}
\omega_1^2=\frac{\left(4/A_2^4\right)\left(u_0\sec\alpha\right)^3<\psi_{0Z}^2>_s}{(1+2\beta^2)<\psi_0^2>_s+\left(2su_0{\beta^2A_4}/{\sqrt{A_2}\cos\alpha}\right)<\psi_0^3>_s},\label{e3}
\end{eqnarray}
\end{widetext}
with
\begin{eqnarray}
<\psi_0^2>_s=\frac{1}{2}\left[H_1\left(1,1\right)-\frac{\pi}{2}\ \beta H_3\left(\frac{3}{2},\frac{3}{2}\right)\right],\label{e4}
\end{eqnarray}

\begin{eqnarray}
<\psi_0^3>_s=\frac12s\beta\left[\frac{\pi}{4}H_1\left(\frac{3}{2},\frac{3}{2}\right)-2\beta H_3\left(2,2\right)\right],\label{e5}
\end{eqnarray}
\begin{eqnarray}
<\psi_{0Z}^2>_s=\frac{1}{6}\left[H_1\left(2,1\right)-\frac{3\pi}{4}\beta H_3\left(\frac{3}{2},\frac{5}{2}\right)\right].\label{e6}
\end{eqnarray}
Also, $\beta=1/\sqrt{1+6\left(A_4/\sqrt{A_2}\right)\cos\alpha}$, $H_n\left(a,b\right)={_2}F_1\left(a,b;\frac{n}{2};\beta^2\right)$ in which the hypergeometric function is represented by  ${_2}F_1$ and the plane solitary pulse solution $\psi_0=\psi(T=0)$.  The profile of the instability growth rate for the mZK soliton is shown in Fig. \ref{fig7-growth-mZK}. It is interesting to note that when the thermal energy of classical electrons is low compared to the rest mass energy $(\beta_{\rm{cl}}\sim0.3)$, the growth rate initially increases until $\alpha$ reaches its value $\alpha\approx1.2$ and then drops down having a cut-off near $\alpha=\pi/2$ (see the solid line). The qualitative feature remains the same with an increasing value of the degeneracy parameter $r_{d0}$ (see the solid and dotted lines). However, as the parameter $\beta_{\rm{cl}}$ is slightly increased from $\beta_{\rm{cl}}\sim0.3$ to $\beta_{\rm{cl}}\sim0.8$, the interval of $\alpha$ in which the growth rate initially increases is reduced from $0\leq\alpha\lesssim1.2$ to $0\leq\alpha\lesssim0.65$ (see the dashed line). In this case, the growth rate tends to reduce in rest of the interval of $\alpha$ with a cut-off at the same value as that for the solid and dotted lines.  
\section{Applications and discussions} \label{sec-applica}
White dwarfs are one of the most studied stellar objects. They have unique physical properties and offer a variety of challenging problems that require proper modeling of their structure, evolution, and radiation spectra, together with advanced theories and experiments and considerable computational efforts. In this respect, new astronomical observations will provide more detailed perspectives of these stars and reveal new phenomena that may challenge our models \cite{saumon2022}. In the present work, we attempt to put forward the existing theory and provide essential guidelines for realizing the localization of small-amplitude electron-acoustic solitons and their stability against transverse plane wave perturbations in high-energy-density plasmas that are relevant, e.g., in the environments of white dwarfs. The results may also apply to other systems like laser-produced plasmas, which we do not discuss here.  
\par 
It has been noted that  non-thermal radiation of relativistic plasma jets is formed in the stellar magnetosphere by accretion-induced collapse of white dwarfs to black holes \cite{kryvdyk2007} in which the stellar magnetic field and the particle density increase considerably. Typically, the accretion flow  strikes the white dwarf photosphere with a free-fall velocity $\approx 5000$ km/s or more, leading to form radiative reverse shocks. As a result, the shock-heated matter gets ionized and heated to temperatures $10-50$ keV  \cite{som2017}. Thus, one can have a high-energy environment of fully ionized gases of highly dense relativistic completely degenerate (at zero-temperature) electrons around white dwarfs that can combine with a smaller fraction of classical hot relativistic accreting astrophysical electron flows to form a two-electron-temperature plasma with a neutralizing stationary ion background immersed in a static magnetic field \cite{shatashvili2020}. We have shown that such plasmas can support the excitation of low-frequency electron-acoustic waves (EAWs), which has been a topic of many researchers in the context of laboratory, space, and astrophysical environments over the last many years (See, e.g., Refs. \cite{vasko2017,dillard2018,pakzad2020,misra2021,ali2023} ).  Such waves in high-energy density plasmas can propagate with a velocity, given by Eq. \eqref{eq-phase},  $v_p\sim2\times10^9$ cm/s (in dimension), which is larger than the particle's free-fall velocity $\approx 5000$ km/s, in the highly relativistic degenerate dense regime with density, $n_{d0}\sim10^{35}~\rm{cm}^{-3}$ for $r_{d0}\sim100$ and temperature, $T_{\rm{cl}}\sim10^8$ K for $\beta_{\rm{cl}}\sim0.1$ (See Fig. \ref{fig1-phase}).             
\par 
At the nonlinear stage, electrostatic perturbations can evolve as slowly varying electron-acoustic solitons in strongly magnetized relativistic plasmas with magnetic field, $B_0\gtrsim10^8$ T. The localization of EAWs can have several significant impacts, e.g., on the temperature-pressure balance in the atmosphere of white dwarfs \cite{valyavin2011}. Also, characteristics of electron-acoustic solitons can help better understanding the internal structures of elementary particles. The evolution of these electron-acoustic solitons is governed by two nonlinear equations, namely the ZK and the mZK equations, defined in two different density regimes \cite{saumon2022,valyavin2011,kryvdyk2007,kryvdyk1999}:  $n_{d0}<7.5\times10^{34}~\rm{cm}^{-3}$ and   $n_{d0}\gtrsim7.5\times10^{34}~\rm{cm}^{-3}$, corresponding to $r_{d0}<50$ and $r_{d0}\gtrsim50$ respectively. The ZK solitons are typically of compressive-type in the regions from moderate to strong relativistic degenerate regimes with density, $n_{d0}\gtrsim10^{31}~\rm{cm}^{-3}$ and temperature, $T_{\rm{cl}}\gtrsim10^8$ K  and rarefactive  in the weakly relativistic regimes  with density, $n_{d0}\sim(10^{28}-10^{31})~\rm{cm}^{-3}$ and temperature, $T_{\rm{cl}}\lesssim10^8$ K (See the discussion in Sec. \ref{sec-appendix}). While the amplitude of compressive ZK solitons can vary in the range $\psi\sim0.1-0.3$, i.e., with electrostatic potential energy, $e\phi\sim(2-6)\times10^{-5}$ J, those of compressive type can have relatively low potential energy, $|e\phi|\sim(2-6)\times10^{-7}$ J. In contrast to rarefactive solitons, the pulse energy of compressive ZK solitons, ${\cal E}$ can be as high  as one enters from weakly to unltra-relativistic degenerate regimes. This is in consistent with the increase in wave amplitude (See subplot (a) of Fig. \ref{fig4-sol-energy}) by the effects of the degeneracy parameter $r_{d0}$.
In the first-order smallness of long-wavelength perturbations, the ZK solitons are found to be unstable while propagating at an angle $\alpha<\alpha_c\equiv37.8^{\circ}$. The growth rate tends to decrease as $\alpha$ approaches $\alpha_c$. The typical growth time of instability can be estimated as $\tau\sim10^{-16}$ s, which remains longer than the oscillation time $\tau_p$ and shorter than the collision time $\tau_c$. However, in the second-order smallness of perturbations, the growth rate tends to increase as $ \alpha$ approaches from $37.8^{\circ}$ to  $90^{\circ}$, and thus the estimated time of instability can be reduced.    
\par 
On the other hand, in ultra-relativistic regimes \cite{saumon2022,valyavin2011,kryvdyk2007,kryvdyk1999,misra2012} with particle density, $n_{d0}\gtrsim7.5\times10^{34}~\rm{cm}^{-3}$ and temperature, $T_{\rm{cl}}\gtrsim10^8$ K, the mZK solitons appear as only compressive type. Their amplitudes ($\psi\sim0.1-0.2$) and energies (${\cal E}\sim5-20$) tend to decrease with increasing values of the degeneracy parameter or the particle density beyond their critical values: $r_{d0}\sim50$ or $n_{d0}\sim7.5\times10^{34}~\rm{cm}^{-3}$. The instability growth rate for mZK solitons can achieve maximum values at an intermediate angle (say, $\alpha_0$) in $0\lesssim\alpha\lesssim90^{\circ}$ for which the typical growth time becomes minimum. However, the growth rate tends to decrease beyond $\alpha_0$ and vanish near $\alpha=90^{\circ}$. Thus, electron-acoustic solitons propagating nearly perpendicular to the external magnetic field in the ultra-relativistic degenerate dense regime are more likely to be stable.    
\par 
From the above discussions, it may be concluded that in the environments of compact astrophysical objects like white dwarfs, the localization of EAWs in relativistic degenerate plasmas with two-temperature electrons is possible and the stability and instability of electron-acoustic solitons under long-wavelength perturbations may be helpful in understanding the onset of soliton turbulence in which significant heating and acceleration of particles occur \cite{song1989}. Furthermore, the soliton instability can be a signature of wave collapse in the multi-dimensional propagation of EAWs in which the wave amplitude grows and the wave energy gets spontaneously concentrated in a small area of space. \cite{kuznetsov1986}. 
To understand how the present investigation of EAWs modifies and advances some previous ones, we consider a recent work by Ali \textit{et al.} \cite{ali2023}. In the latter, the authors studied the one-dimensional propagation of large-amplitude EAWs in plasmas with cool and hot degenerate electrons but without any magnetic field effects. In their study, they did not consider any relativistic motion of fluid particles  and study the stability of solitons. Furthermore,  they  reported the existence of only rarefactive solitons. Surprisingly, they observed the appearance of some shocklet structures by the effects of electron degeneracy  and the number density ratio. The appearance of such structures may be due to some incorrect assumptions of the equation of state, which may not be applicable for plasmas with finite-temperature degenerate electrons, and that the hot populations are denser than cool populations. In this respect, the present model is more consistent, and the results obtained can better fit with astrophysical observations to be available in the near future.  In another study \cite{misra2021},  the effects of Landau damping due to multi-plasmon resonances on electron-acoustic solitons were studied using the kinetic theory approach in a two-electron-temperature plasma with finite temperature degeneracy. The characteristics of the phase velocity of EAWs reported there are similar to those presented in the present work.  
 \section{Summary and conclusion} \label{sec-summary}
We have studied the nonlinear propagation of small-amplitude EAWs in a magnetoplasma with a relativistic flow of sparsely populated classical thermal electrons and relativistic fully degenerate dense electrons (main constituents) immersed in a neutralizing background of positive ions. Using the Lorentz transformations for the space and time coordinates and the standard reductive perturbation technique, we have derived the ZK and mZK equations for the evolution of electron-acoustic solitons in two different regimes of relativistic degeneracy, namely $r_{d0}<50$ and $r_{d0}\gtrsim50$, i.e., in the regimes of weakly to strongly relativistic degenerate plasmas and strongly to ultra-relativistic plasmas. In both cases, we have obtained traveling wave plane soliton solutions and the expressions for the soliton energies. We have shown that electron-acoustic solitons can propagate with a speed larger than the linear phase speed of EAWs, leading to positive dispersion at which longer wavelength perturbations travel faster than those with shorter wavelengths. Furthermore, the stability of these solitons under the long-wavelength transverse plane wave perturbations is studied using the small-$k$ expansion method \cite{allen1995}. The main findings of this investigation are summarized as follows:  
\begin{itemize}
\item Low-frequency (in comparison with the plasma frequency of classical electrons) long-wavelength small-amplitude EAWs (linear mode) can propagate in relativistic degenerate magnetoplasmas with phase velocity $(v_p)$ smaller than the speed of light in vacuum $(c)$. 
\item In contrast to typical classical plasmas with hot and cold electron species \cite{vasko2017,dillard2018} or plasmas with finite temperature degeneracy of two-temperature electrons \citep{misra2021},  the linear phase velocity of EAWs remains smaller than the thermal velocity of classical electrons in the regime of weakly relativistic temperature ($k_BT\ll m_ec^2$ or $\beta_{\rm{cl}}\ll1$) and weakly relativistic degeneracy ($p_F\ll m_e c$ or $r_{d0}\ll1$) of electrons. In this case, the phase velocity reduces to $v^{\rm{cl}}_p\approx\sqrt{\beta_{\rm{cl}}/G_{\rm{cl}}}$, i.e., the classical reult is recovered.  
\item The phase velocity tends to increase in the regime of moderate relativistic degeneracy, i.e., $0<r_{d0}<10$ and  weakly relativistic temperature $\beta_{\rm{cl}}\ll1$. However, in the regimes of moderate to ultra-relativistic temperatures, it tends to decrease with $r_{d0}$ but approaches a constant value in the ultra-relativistic degeneracy regime $(r_{d0}\gg1)$. The phase velocity is enhanced significantly with an increase of the parameter $\beta_{\rm{cl}}$ from its value smaller to grater than unity. In the limit of  $r_{d0}\gg1$, we have $v^{\rm{ur}}_p\approx r_{d0}/\sqrt{3R_0^3}$.
\item Depending on the parameter regimes we choose for $\beta_{\rm{cl}}$ and $r_{d0}$, the EAWs can propagate as the ZK or mZK solitary waves and they can evolve in the form of either compressive type (with positive potential) or rarefactive type (with negative potential) with speeds larger than the phase speed of linear EAWs. While the compressive solitons are more likely to appear in higher density regimes ($n_{d0}\gtrsim10^{31}~\rm{cm}^{-3}$) with higher temperature ($T_{\rm{cl}}\gtrsim10^8$ K), the rarefative solitons occur in relatively low temperature regimes ($\lesssim10^8$ K) with density  $7\times10^{28}\lesssim n_{d0}\lesssim4\times10^{31}~\rm{cm}^{-3}$.
Also, ZK solitons can propagate with increasing energies as one enters from weakly to ultra-relativistic regimes of thermal and Fermi energies of electrons, i.e.,  with increasing values of both the parameters $\beta_{\rm{cl}}$ and $r_{d0}$. 
\item For $r_{d0}\gtrsim50$, which corresponds to the density regime, $n_{d0}\gtrsim10^{34}~\rm{cm}^{-3}$, the nonlinear coefficient $(A_1)$ of the ZK equation tends to vanish. In this regime, the evolution of electron-acoustic solitons can not be described by the ZK equation but the mZK equation with an additional cubic nonlinearity. Depending on whether $A_1$ is positive or negative, the mZK solitons can be compressive or rarefactive type. Although the effect of $\beta_{\rm{cl}}$ on the profile of the mZK soliton is similar to the ZK soliton,  both the amplitude and width of the mZK soliton decrease (in contrast to the ZK soliton)  with increasing values of the degeneracy parameter $r_{d0}$. Also, in contrast to the ZK soliton,  the energy of the mZK soliton tend to decrease with increasing values of $r_{d0}$ and the decrement is significant with a small enhancement of $\beta_{\rm{cl}}$.  
\item Obliquely propagating ZK and mZK solitons under transverse plane wave perturbations exhibit instability. However, even if the ZK soliton is stable for the first-order perturbations, it can be unstable at the second-order smallness of wave frequency or wave number. While the first-order growth rate of instability of ZK solitons tends to decrease with the angle of propagation $\alpha$ having cut-off at $\alpha\approx 38^\circ$, the second-order growth rate increases with $\alpha\gtrsim38^\circ$. Both the first and second-order growth rates get enhanced with an increase of the degeneracy parameter $r_{d0}$ but they decrease with an increment of the relativity parameter $\beta_{\rm{cl}}$. On the other hand, only the first-order growth rate of instability for the mZK soliton is obtained. It increases with $\alpha$, reaches its maximum value, and then decreases having a cut-off at $\alpha\approx\pi/2$. The growth rate always increases with an increase of $r_{d0}$. However, depending on $\alpha$, the same can be increased or decreased with an increase of $\beta_{\rm{cl}}$. 
\end{itemize}
\par 
To conclude, the results obtained in this work will help understand the localization of two-dimensional EAWs in relativistic degenerate magnetoplasmas, such as those in the environments of compact astrophysical bodies like white dwarfs, which consist of a core of degenerate electrons providing the bulk of the mass, surrounded by a thin non-degenerate gas of low-density electrons and stationary ions. The stability and instability of ZK solitons at the first and second-order perturbations can help realize soliton turbulence in nonlinear media similar to soliton Langmuir turbulence in which significant heating and acceleration of particles can occur \cite{song1989}. Since the two-dimensional ZK and mZK solitons exhibit instability under a transverse plane wave perturbation, such instability can be a signature of wave collapse \cite{kuznetsov1986}. In the latter, the wave amplitude grows exceedingly high at a point after a finite time, and the wave energy gets spontaneously concentrated in a small area of space with its subsequent dissipation. However, the detailed analysis of the collapse theory is beyond the scope of the present work. Finally, the electron-acoustic solitary pulses should play a vital role in the transportation and localization of energy. They may also act as a route to the fragmentation of giant gas clouds into dense cores.
\section*{Author Contributions}
\textbf{Amar Prasad
Misra:} Model proposal; Conceptualization; Formal analysis (equal); Investigation (equal); Methodology (equal); Writing-original draft (equal), Review \& editing;  Validation (equal).
 \textbf{Alireza Abdikian:} Formal analysis (equal); Investigation (equal);
Methodology (equal); Writing-original draft (equal), validation (equal). 
\section*{Conflict of interest} The authors have no conflicts to disclose.
 \section*{Data availability statement} 
 All data that support the findings of this study are included within the article (and any supplementary files).
\appendix
\section{Derivation of the relativistic fluid model} \label{sec-appendix}
Here, we give some details of the derivation of the basic fluid model as in Sec. \ref{sec-model}.
\par 
We consider the following energy-momentum tensor for an ideal fluid.
\begin{equation}
T^{\mu\nu}=\frac{H}{c^2}u^{\mu}u^{\nu}-Pg^{\mu\nu},
\end{equation}
where $g^{\mu\nu}=\rm{diag}(1,-1,-1,-1)$ is the Minkowski metric tensor, $H$ is the enthalpy per unit volume, $P$ is the fluid pressure, $c$ is the speed of light in vacuum, and $u^\mu=\gamma\left(c,{\bf v}\right)=\gamma\left(c,v^l\right),~l=1,2,3$, is the average fluid velocity with $\gamma=1/\sqrt{1-v^2/c^2}$ denoting the relativistic gamma factor. Furthermore, while $\mu,~\nu=0$  correspond to the time, the values $\mu,~\nu=1,2,3$ correspond to the space components. For the metric tensor we also use the signature: $ds^2=g_{\mu\nu}dx^\mu dx^\nu$ with $dx^\mu=\left(cdt,dx^l\right),~l=1,2,3$. We define $n$ as the particle number density in the proper frame so that $N=\gamma n$ that in the laboratory frame and ${\cal E}$ as the total energy density such that $H={\cal E}+P$. 
\par 
In the case of a completely degenerate plasma, $nk_BT/P\ll1$ for which one can use the following pressure law for a zero-temperature Fermi gas \cite{chandrasekhar1935}:
\begin{equation}
P=\frac{m_e^4c^5}{24\pi^2\hbar^3}\left[r_d\left(2r_d^2-3\right)\left(1+r_d^2\right)^{1/2}+3\sinh^{-1}r_d\right]. \label{eq-press}
\end{equation}
Considering the weakly-relativistic $(r_d\ll1)$ and ultra-relativistic $(r_d\gg1)$ degeneracy limits, the pressure equation \eqref{eq-press} can be rewritten in the following form \cite{mikaberidze2015}:
 \begin{equation}
 \frac{P}{P_0}=\left(\frac{n}{n_0}\right)^\Gamma \label{eq-press-deg}
 \end{equation}
with a polytropic index $\Gamma$: $4/3\leq\Gamma\leq5/3$, where $\Gamma=5/3$ for the weakly relativistic or nonrelativistic degenerate electrons $(r_d\ll1)$ and $\Gamma=4/3$ for ultra-relativistic degeneracy $(r_d\gg1)$. The particle distribution function remains locally Juttner-Fermian. Also, for zero-temperature Fermi gases, all the quantities ${\cal E}$, $H$ and $P$ depend on the number density $n$ only. The system is isentropic and also, as the temperature tends to zero, the entropy approaches zero too \cite{mikaberidze2015}.  
\par 
On the other hand, for classical nondegenerate relativistic plasmas, we can also consider the pressure law in the same form as Eq. \eqref{eq-press-deg} \cite{gratton1997,lee2007}:
\begin{equation}
 \frac{P}{P_0}=\left(\frac{n}{n_0}\right)^\Gamma \label{eq-press-cl}
 \end{equation}
with a polytropic index $\Gamma$: $4/3\leq\Gamma\leq5/3$, where $\Gamma=5/3$ for the weakly relativistic or nonrelativistic thermal motion of electrons $(P\ll nmc^2)$ and $\Gamma=5/3$ for high-temperature or ultra-relativistic fluid flow $(P\gg nmc^2)$. 
\par
The conservation of number of particles gives
\begin{equation}
\partial_\nu\left(nu^\nu\right)=0. \label{eq-conser-num}
\end{equation}
The conservation of momentum gives (in presence of electromagnetic fields ${\bf E}$ and ${\bf B}$)
\begin{equation}
\partial_\nu T^{\mu\nu}=\frac{q}{c}N_\nu F^{\nu\mu}, \label{eq-conser-mom}
\end{equation}
where $N^\nu=nu^\nu$ is the particle flux (four-vector), $q$ is the electric charge, and the force $F^{\nu\mu}=[{\bf E},~{\bf B}]$ is defined as \cite{lee2007}  
\begin{equation}
F^{oi}=E_i,~F^{ij}=\epsilon_{ijk}B_k.
\end{equation}
Next, using Eq. \eqref{eq-conser-num} and noting that $u^\mu u_\mu=c^2$, the equation $u_\mu\partial_\mu T^{\mu\nu}=0$ for $\mu=\nu=0$ reduces to the following equation.
\begin{equation}
\partial_\nu\left(\frac{H}{n}\right)=\frac{1}{n}\partial_\nu P,
\end{equation}
which gives \cite{gratton1997,lee2007,son2006}  
\begin{equation}
\frac{d}{dt}\left(\frac{H}{n}\right)=\frac{1}{n}\frac{dP}{dt}. \label{eq-HP}
\end{equation}
Next, using the relation $\left(P/P_0\right)=\left(n/n_0\right)^\Gamma$, we obtain from Eq. \eqref{eq-HP} the following relations \cite{gratton1997,lee2007,son2006}. 
\begin{equation}
{\cal E}=\frac{P}{\Gamma-1}+nmc^2,
\end{equation}
\begin{equation}
H=\left(\frac{\Gamma}{\Gamma-1}\right)P+nmc^2;~4/3\leq\Gamma\leq5/3,
\end{equation}
where we have used the condition that at $P=0$, ${\cal E}=n_0mc^2$. 
\par  
Equation \eqref{eq-conser-num}, after using $u^0=\gamma c$ and $\partial_0=(1/c)(\partial/\partial t)$, gives the following equation of continuity.
\begin{equation}
\partial_0\left(nu^0\right)+\partial_j\left(nu^j\right)=0,\rm{i.e.,}~\frac{\partial}{\partial t}(\gamma n)+\nabla\cdot (\gamma n{\bf v})=0.\label{eq-contA}
\end{equation}
We note that $T^{0\mu}=T^{\mu0}=(H/c^2)\gamma^2v^\mu$, $\mu=1,2,3$. Thus, Eq. \eqref{eq-conser-mom} reduces to
\begin{equation}
\partial_0T^{\mu0}+\partial_\nu T^{\mu\nu}=\frac{q}{c}nu_\nu F^{\nu\mu};~\mu,\nu=1,2,3,
\end{equation}
or, we have
\begin{equation}
\begin{split}
\partial^\mu P&+\left(n u^\nu\right)\partial_\nu\left(\frac{H}{nc^2}u^\mu\right)\\
&+\frac{H}{nc^2}u^\mu\partial_\nu\left(n u^\nu\right)=\frac{q}{c}nu_\nu F^{\nu\mu}-\frac{1}{c}\partial_0\left(H\gamma u^\mu\right).
\end{split}
\end{equation}
Simplifying further, we get
\begin{equation}
\left(n u^\nu\right)\partial_\nu\left(\frac{H}{nc^2}u^\mu\right)+\gamma n \partial_t\left(\frac{H}{nc^2}  u^\mu\right)+\partial^\mu P=\frac{q}{c}nu_\nu F^{\nu\mu}. \label{eq-mm}
\end{equation}
Dividing throughout by $\gamma n$ and noting that $u^\nu=\gamma v^\nu$, $\nu=1,2,3$, we obtain (setting $\mu=i,~\nu=j$) from Eq. \eqref{eq-mm} the following \cite{lee2007}.
\begin{equation}
\begin{split}
\left(\frac{\partial}{\partial t}+v_j\frac{\partial}{\partial x_j}\right)&\left(\frac{H}{nc^2}\gamma v_i\right)+\frac{1}{\gamma n}\frac{\partial P}{\partial x_i}=\frac{q}{c}v_j F^{ji}\\
&\equiv q\left[E_i+\frac{1}{c}\left(v\times B\right)_i\right]. \label{eq-mm2}
\end{split}
\end{equation}
In vector form, we rewrite Eq. \eqref{eq-mm2} as \cite{lee2007} 
\begin{equation}
\frac{1}{c^2}\left(\frac{\partial}{\partial t}+{\bf v}\cdot\nabla\right)\left(\frac{H\gamma {\bf v}}{n}\right)+\frac{1}{\gamma n}\nabla P=q\left({\bf E}+\frac{1}{c} {\bf v}\times {\bf B}\right).\label{eq1}
\end{equation}
Next, defining the time scales for the density and pressure fluctuations, respectively, by $\tau_n$ and $\tau_p$, where $\tau_p^{-1}=\left(1/P\right)\left(dP/dt\right)$ and $\tau_n^{-1}=\left(1/n\right)\left(dn/dt\right)$, the total derivative by $d/dt\equiv \partial_t+{\bf v}\cdot\nabla$, and using the relation $H=nmc^2+\left[\Gamma/(\Gamma-1)\right]P$, we recast Eq. \eqref{eq1} as
\begin{equation}
\begin{split}
\frac{H}{nc^2}\frac{d}{dt}(\gamma {\bf v})&+\frac{\gamma {\bf v}}{c^2}\frac{\Gamma}{\Gamma-1}\frac{P}{n}\left(\tau_p^{-1}-\tau_n^{-1}\right)=-\frac{1}{\gamma n}\nabla P\\
&+q\left({\bf E}+\frac{1}{c} {\bf v}\times {\bf B}\right). \label{eq-mm3}
\end{split}
\end{equation}
The time variation  of electron fluid pressure is small compared to that of the density fluctuations. This is valid for low-frequency acoustic-like perturbations having phase velocity lower than the speed of light $c$ (See, e.g., the dispersion relation \eqref{eq-phase} in the text and Ref. \cite{haas2016}). Thus, one has $\tau_n\gg\tau_p$ and after multiplying throughout by $\gamma n$, we have from Eq. \eqref{eq-mm3} the following \cite{gratton1997,misra2018}.  
\begin{equation}
\frac{\gamma H}{c^2}\frac{d}{dt}(\gamma {\bf v})= q\gamma n\left({\bf E}+\frac{1}{c} {\bf v}\times {\bf B}\right)-\left(\nabla+\frac{\gamma^2{\bf v}}{c^2}\frac{d}{dt}\right)P. \label{eq-moment}  
\end{equation}
Finally, we consider the component corresponding to $\mu=0$ of the Gauss law:
\begin{equation}
\partial_\nu F^{\mu\nu}=\frac{4\pi}{c}\sum_jq_jN^\mu_j 
\end{equation} 
to obtain the following Poisson equation.
\begin{equation}
 \nabla^2\phi=-4\pi\sum_jq_j\gamma_jn_j.\label{eq-poissA}
\end{equation}
Equations \eqref{eq-contA}, \eqref{eq-moment}, and \eqref{eq-poissA} constitute the required fluid model as in Sec. \ref{sec-model}, which exactly agree with those in Ref. \cite{misra2018}. They also agree with those for relativistic degenerate fluids as in Refs. \cite{kerr2014,haas2016}  after a minor correction with the $\gamma$-factor is made in the momentum balance equations therein.
\bibliographystyle{apsrev4-1}
\bibliography{ref}
\nopagebreak


\end{document}